\documentclass[useAMS,usenatbib]{mn2e}
\usepackage{graphicx,mathptm,amsmath,amssymb}

\title[MEM image reconstruction using wavelets]
  {Maximum-entropy image reconstruction using wavelets}
\author[Klaus Maisinger, M.P.~Hobson and A.N.~Lasenby]
{Klaus Maisinger, M.P.~Hobson and A.N.~Lasenby\\
Astrophysics Group, Cavendish Laboratory, 
Madingley Road, Cambridge, CB3 0HE, UK}

\date{Accepted ---. Received ---; in original form 9th March 2003}

\pagerange{\pageref{firstpage}--\pageref{lastpage}} \pubyear{2003}

\newcommand{\myvec}[1]{{\mathbfit{#1}}}
\newcommand{\mymatrix}[1]{{\mathbfss{#1}}}
\newcommand{\mathset}[1]{\mathbb{#1}}

\def\sspd#1#2{{\frac{\upartial ^2 #1}{\upartial #2^2}}}
\def\rms{\textit{rms}}
\def\transp{\mathrm{t}}

\def\reff@jnl#1{{\rm#1\/}}
\def\aap{\reff@jnl{A\&A}}               
\def\aaps{\reff@jnl{A\&AS}}             
\def\mnras{\reff@jnl{MNRAS}}            
\def\apjl{\reff@jnl{ApJ}}               

\begin{document}

\label{firstpage}

\maketitle

\begin{abstract}
  Wavelet functions allow the sparse and efficient representation of a
  signal at different scales. Recently the application of wavelets to
  the denoising of maps of cosmic microwave background (CMB)
  fluctuations has been proposed.  The maximum-entropy method (MEM) is
  also often used for enhancing astronomical images and has been
  applied to CMB data.  In this paper, we give a systematic discussion
  of combining these two approaches by the use of the MEM in wavelet
  bases for the denoising and deconvolution of CMB maps and more
  general images.  Certain types of wavelet transforms, such as the
  \`{a} trous transform, can be viewed as a multi-channel intrinsic
  correlation function (ICF).  We find that the wavelet MEM has lower
  reconstruction residuals than conventional pixel-basis MEM in the
  case when the signal-to-noise ratio is low and the point spread
  function narrow.  Furthermore, the Bayesian evidence for the wavelet
  MEM reconstructions is generally higher for a wide range of images.
  From a Bayesian point of view, the wavelet basis thus provides a
  better model of the image.
\end{abstract}


\section{Introduction}

Both the maximum-entropy method (MEM) and wavelet techniques are used
for astronomical image enhancement.  In particular, both methods have
recently been applied to the analysis of CMB data (see, for instance,
\citealt*{hobson99}; \citealt{sanz99a,sanz99b,tenorio99}). Maps of CMB
anisotropies are a useful tool in the analysis of CMB data.  Making
maps is rarely straightforward, since a multitude of systematic
instrumental effects, calibration uncertainties and other deficiencies
in the modelling of the telescope come into play.  For example,
interferometric maps suffer from the telescope's incomplete sampling
in Fourier space and require the deconvolution of the synthesised beam
\citep*[e.g.][]{thompson94} and the suppression of receiver noise.  CMB
observations from single-dish telescopes use total power measurements
and scan across the observed fields to assemble a map.  Here it is the
effect of the finite primary beam that needs to be deconvolved in
order that a high-resolution map may be recovered.  Beyond the area of
the CMB, the task of image reconstruction is generic and occurs in
virtually any type of astronomical map-making.

In a general imaging problem, we assume that the data~$d$ observed by
an experiment are given by a convolution of the true sky signal, or
image~$h$, with the point spread function~$P$ of the instrument, plus
some Gaussian random noise~$n$:
\[
d = P \ast h + n.  
\]
In the discretised version, the data vector~$\myvec{d}$ is given by a
multiplication of the vector~$\myvec{h}$ of the image pixels with the
instrumental response matrix~$\mymatrix{R}$ that describes the
convolution with the point spread function, and the additive noise
vector~$\myvec{n}$:
\[
\myvec{d} = \mymatrix{R} \myvec{h} + \myvec{n}.
\]
To solve the inverse problem of recovering the image~$\myvec{h}$ from
the data, some type of regularisation is usually required.  A common
technique is to use an entropic function~$S$ for the
regularisation.  The best reconstruction is then found by minimising
the function~$F(\myvec{h}) = \tfrac{1}{2} \chi^2(\myvec{h}) - \alpha
S(\myvec{h})$ that determines a suitable trade-off between a good fit
to the data enforced by the $\chi^2$-statistic and a strong
regularisation given by the entropy~$S(\myvec{h})$ of the
reconstruction.  The maximum-entropy method has proven to be very
successful for the deconvolution of noisy images.

Despite its capabilities, the MEM suffers from several shortcomings.
For example, the appropriate entropy functional depends on the
properties of the distribution of image pixels, but it is not always
evident what the theoretical distribution should be.  For positive
additive distributions, one uses the entropy
\begin{equation}
S(\myvec{h}) = \sum_{i=1}^{N_h} h_i - m_i - h_i \log \frac{h_i}{m_i},
\label{eqn:gullskilling}
\end{equation}
where the sum is over all image pixels and $m_i$ is a measure assigned
to pixel~$i$.  Even in this case problems can arise when there is no
appropriate background level, or the image brightness falls below the
background level in some areas.

Another defect of the MEM is that, in its simplest forms, correlations
between image pixels are not taken into account properly.  This
problem manifests itself in several guises.  Because of correlations
between image pixels, the effective number of `degrees of freedom' in
the data is often much smaller than the number of parameters in the
minimisation problem, making effective regularisation more difficult.
In fact, MEM is inherently based on the assumption that image pixels
are independent.  Furthermore, ignoring correlations leads to the
introduction of spurious features in the map, such as the
characteristic ringing present on uniform backgrounds.  There is no
provision in the MEM algorithm to reward local smoothness of the
image.  It appears to be quite difficult to regularise in such a way
as to reconstruct faithfully sharp features and uniform areas at the
same time.

Several solutions have been proposed to remedy the problem of image
correlations.  \cite{gullskilling99} have introduced the concept of an
intrinsic correlation function (ICF) that is used to decorrelate the
reconstructed image.  The ICF framework has been extended to allow
reconstructions of objects on different scales. \citet{weir92}
proposes a multi-channel approach, which allows for multiple scales of
pixel-to-pixel correlations.  In pyramidal maximum entropy
\citep*{bontekoe94}, the number of pixels retained in the
low-resolution channels is decimated.  Despite these improvements,
choosing an ICF is not straightforward.  It is clear that there is no
single set of ICFs that is universally optimal for all possible types
of data.  Choosing suitable scale lengths and weights is of great
importance.

A slightly different approach to tackling the correlation problem is
to use a representation of the image that is more efficient in
identifying its information content.  In other words, the task is to
find an optimal {\em basis set} for the representation of the image.
Furthermore, it is desirable to have a representation that can
efficiently capture information present on different length scales in
the image.  For instance, for CMB observations several foreground
components, such as radio point sources or SZ-clusters, are very
localised on the sky.  Some theories for structure formation also
predict localised non-Gaussian imprints at arcminute scales
on the CMB itself, for example
temperature fluctuations produced in the wake of cosmic strings.  On
the other hand, the primordial CMB itself shows more diffuse structure 
that peaks on angular scales close to a degree. 
Representing these signals in real
space (i.e. the image plane) requires large numbers of basis functions
(i.e. the pixels) for a given image.  Similarly, a reconstruction in
Fourier space requires the determination of a large number of modes,
which are often very poorly constrained by the data, since each
localised feature on a map is expanded into an infinite number of
basis functions in Fourier space.  Maximum entropy has been applied to
reconstructions in both real and Fourier space.  A Fourier space
approach has been developed by \citet{hobson98} and has been applied
to simulated Planck data, while \citet*{jones99} simulate its use for
the MAP satellite mission.

The application of {\em wavelets} to CMB data analysis has recently
been investigated \citep[see, for
instance,][]{hobson99,sanz99a,sanz99b,tenorio99,cayon00,vielva01}.
Wavelets are special sets of functions that allow the efficient
representation of signals both in real and in Fourier space.
Furthermore, they can represent different objects of greatly varying
sizes simultaneously.  The term `wavelet' does not refer to a single
unique function.  Instead, it comprises a whole class of functions
with similar properties.  In the context of CMB analysis, wavelets
have predominantly been used for noise filtering or the separation of
localised foreground sources. A combination of MEM and certain types
of wavelets has been discussed by \citet{pantinstarck96} and
\citet{starck01}.

In this paper, we investigate the use of wavelets in MEM more closely.
We will compare different wavelet transforms, entropic priors and
regularisations.  We will also consider how the different approaches
can be viewed in the ICF framework.  In
Section~\ref{sec:memwavelet:wavelets} we give an introduction to
wavelet transforms, and the application of wavelet filters to
denoising is reviewed in Section~\ref{sec:memwavelet:filter}. In
Section~\ref{sec:bayes} the maximum-entropy method is introduced. The
combination of wavelet and MEM techniques is explored in
Section~\ref{sec:memwavelet:transforms}, and the equivalence of
intrinsic correlation functions and certain redundant wavelet
transforms is discussed in Section~\ref{sec:memwavelet:multichannel}.
In Section~\ref{sec:memwavelet:application}, we test the techniques presented
by applying them to simulated image reconstruction problems.

\section{The wavelet transform}
\label{sec:memwavelet:wavelets}

Wavelets are functions that enjoy certain properties, which will be
further discussed below.  A wavelet basis can be constructed
from dilations and translations of a given wavelet function.  The
wavelet transform is an integral transform that uses the wavelet basis
functions.  The most widely used classes of wavelet transforms are
orthogonal transforms.  Like the Fourier transform, they are
essentially rotations in function space from the pixel basis to
appropriate wavelet basis functions.  Unlike Fourier transforms, the
basis functions can be well-localised in both real and Fourier space
(even though they can only be compactly supported in one of the two).

\subsection{The continuous wavelet transform}

The continuous wavelet transform of a one-dimensional square
integrable function $f \in \mathcal{L}^2(\mathset{R})$ can be defined
as
\[
W(a,b) =  \int_{-\infty}^{\infty} f(x) \frac{1}{\sqrt{a}} \psi
\left(\frac{x-b}{a}\right) \ \mathrm{d}x,
\]
where the function $\psi(x)$ is the wavelet (often called analysing
wavelet or mother wavelet). The real numbers~$a>0$ and~$b$ are scale
and position parameters respectively.  Dilations and translations
of~$\psi(x)$ can be derived by varying~$a$ and~$b$.  Obviously, the
wavelet transform is linear.  The inverse wavelet transform is given by
\[
f(x) = \frac{1}{C_\psi} \int_0^\infty \!a^{-2} \mathrm{d}a
\int_{-\infty}^\infty \!\mathrm{d}b\ W(a,b)
\frac{1}{\sqrt{a}} \psi
\left( \frac{x-b}{a} \right),
\]
where the normalisation constant $C_\psi = 2 \pi \int_{0}^\infty
|k|^{-1} | \widetilde{\psi} (k)|^2\ \mathrm{d}k$ can be obtained from
the Fourier transform $\widetilde{\psi}(k)$ of the wavelet
function~$\psi(x)$ \footnote{The normalisation of $C_\psi$ assumes a
  Fourier transform that is scaled by $1/\sqrt{2\pi}$ in both directions.}. 

In order to construct a basis from the translations and dilations 
of the wavelet function~$\psi(x)$, a
second function~$\phi(x)$ must be introduced.  It is called the
scaling function or father wavelet, and its relation to~$\psi(x)$ will
become clear in a moment.  For the resulting basis to be discrete, compact and
orthogonal, the wavelet functions~$\psi(x)$ and~$\phi(x)$ must obey a set of
mathematical restrictions first derived by Daubechies
\citep[e.g.][]{daubechies92}, among them
\begin{equation}
\int_{-\infty}^\infty \phi (x)\ \mathrm{d}x= 1,\quad
\int_{-\infty}^\infty \psi (x)\ \mathrm{d}x= 0,
\label{eqn:daubcond1}
\end{equation}
and the normalisation 
\begin{equation}
\int_{-\infty}^\infty |\psi|^2 (x)\  \mathrm{d}x = 1.
\label{eqn:daubcond2}
\end{equation}
A wavelet basis can then be constructed as follows.  It is convenient to
take special values for $a$ and $b$ in defining the basis: $a =
2^{-j}$ and $b = 2^j l$, where $j$ and $l$ are integers labelling the
scale of the wavelet and the position at this scale.  The resulting
wavelet functions are
\begin{eqnarray*}
\phi_{j,l} ( x )& = &2^\frac{j}{2} \phi ( 2^j x - l),\\
\psi_{j,l} ( x )& = &2^\frac{j}{2} \psi ( 2^j x - l).
\end{eqnarray*}
It can be shown that the set $ \{ \phi_{0,l}, \psi_{j,l} \}$ with 
$j \ge 0$ and $-\infty<l<\infty$ forms a complete orthonormal basis in
$\mathcal{L}^2(\mathset{R})$.  One may then expand a function~$f(x)$ as
\begin{equation}
f(x) = \sum_{l=-\infty}^\infty c_{0,l} \phi_{0,l}(x) +
\sum_{j=0}^\infty \sum_{l=-\infty}^\infty w_{j,l}
\psi_{j,l}(x),\label{eqn:contwtexpansion}
\end{equation}
where the wavelet coefficients $c_{0,l}$ and $w_{j,l}$ are given by
\begin{eqnarray}
c_{0,l}& =& \int_{-\infty}^\infty f(x) \phi_{0,l}(x)\ \mathrm{d}x,\nonumber\\
w_{j,l}& =& \int_{-\infty}^\infty f(x) \psi_{j,l}(x)\ \mathrm{d}x.
\label{eqn:contwtcoeff}
\end{eqnarray}

\begin{figure}
  \begin{center}
    \leavevmode
    \begin{tabular}{ll}
      \includegraphics[width=3.8cm]{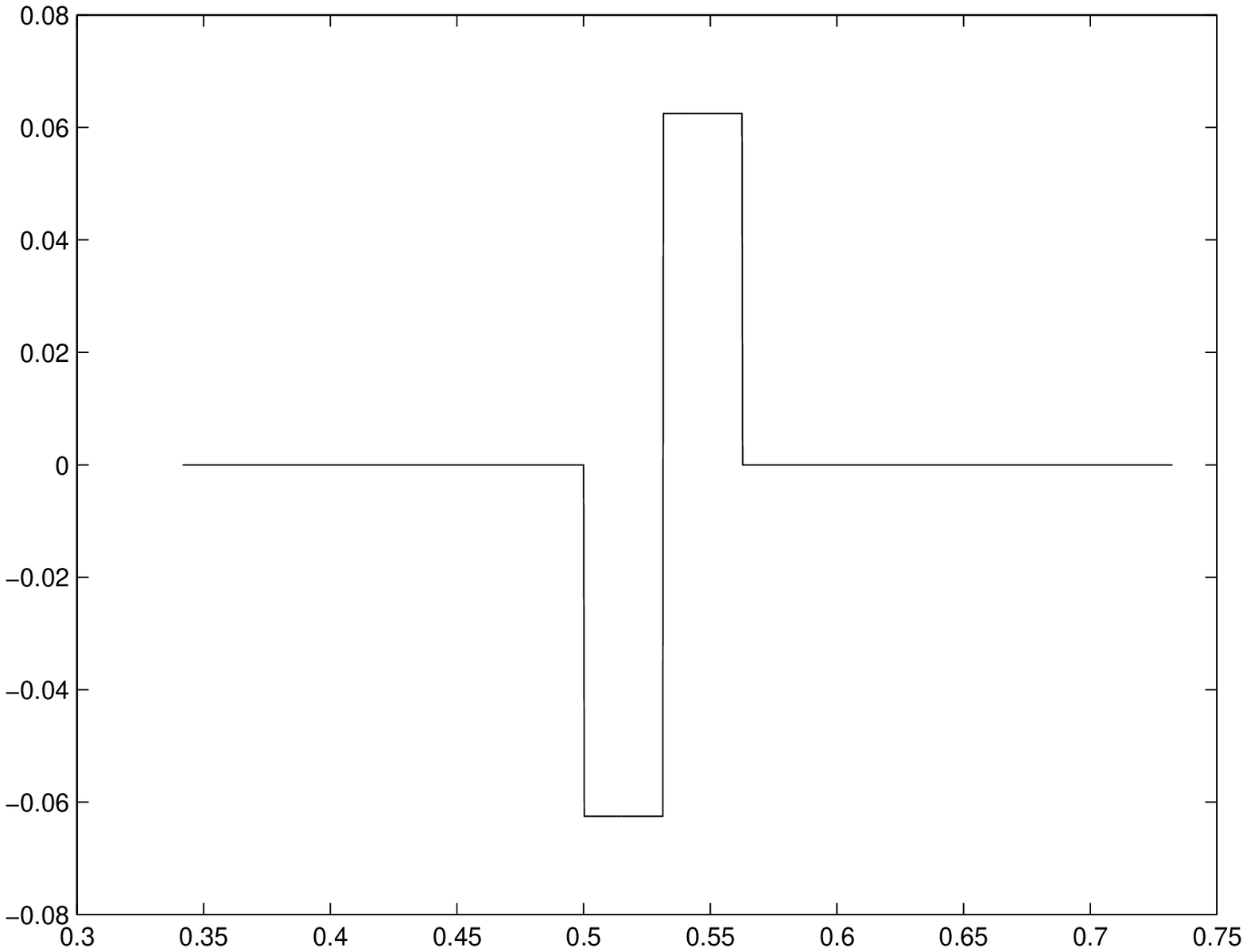}&
      \includegraphics[width=3.8cm]{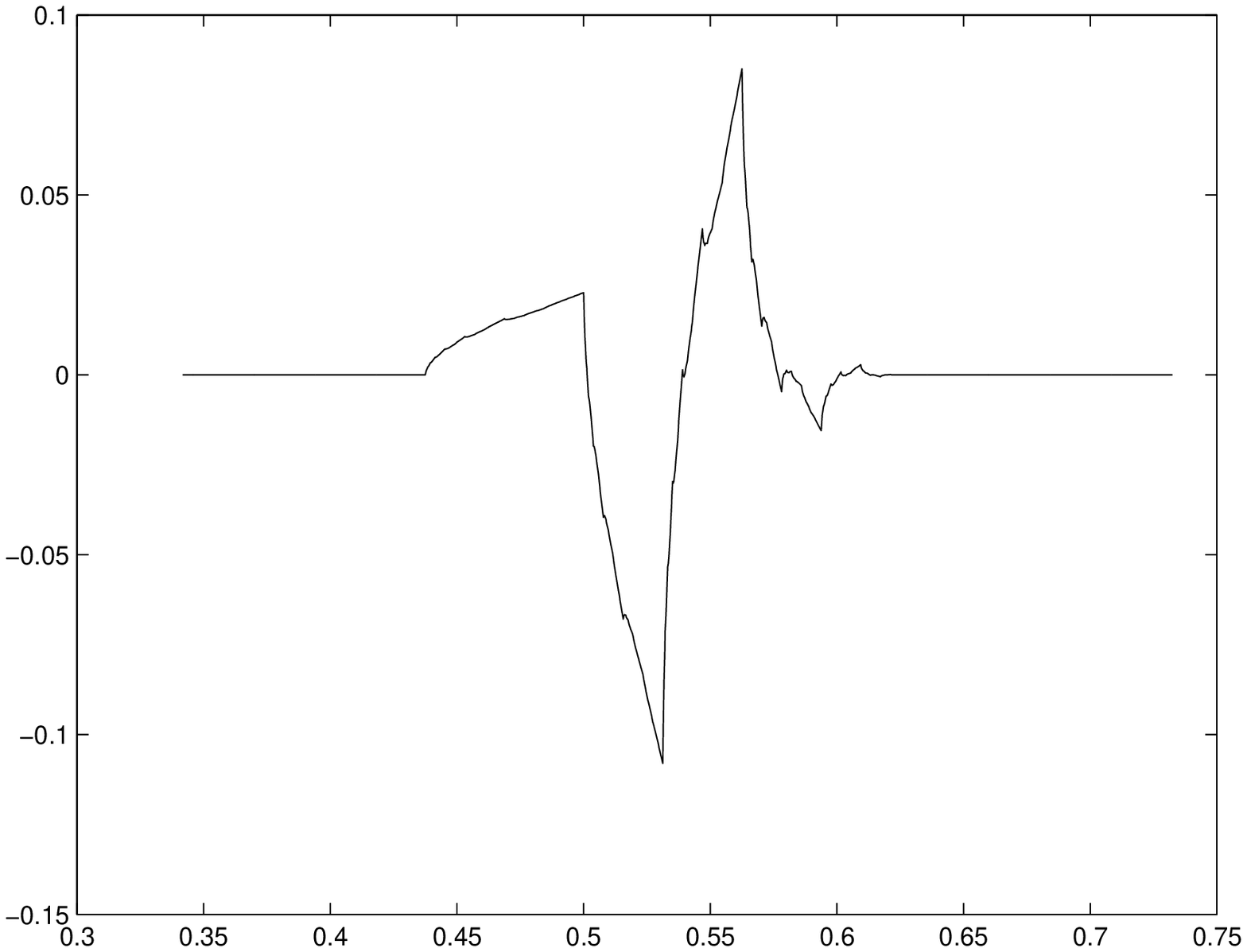}\\
      (a) & (b)
    \end{tabular}
    \caption{Wavelet functions~$\psi(x)$: (a) Haar wavelet and (b)
      Daubechies-4 wavelet.}
    \label{fig:haardaub}
  \end{center}
\end{figure}

Since the wavelet basis consists of dilations and translations of the
mother and father wavelets~$\psi(x)$ and~$\phi(x)$, one can obtain a
different orthogonal 
wavelet basis for each pair of these functions that obey the
Daubechies conditions.  Thus there exists an infinite number of
possible wavelet transforms, with different wavelet bases making
different trade-offs between how compactly they are localised in
either real space or frequency space.  Unfortunately, in nearly all
cases, the wavelet basis functions cannot be expressed in closed form
in terms of simple functions.  Nevertheless, two of the more commonly
used wavelet basis functions, the Haar and Daubechies wavelets, are
plotted in Fig.~\ref{fig:haardaub}.  The properties of wavelet and
scaling functions can be more easily understood within the framework
of multiresolution analysis (MRA), which is discussed in more detail
in Appendix~\ref{sec:wavelet:mra}.

\subsection{The discrete wavelet transform}
\label{sec:wavelet:dwt}

In many applications, one is not so much interested in a
function~$f(x)$ defined at all values of~$x$, as in its samples
$f(x_i)$ at $N=2^J$ equally-spaced points $x_i$.  By analogy with the
discrete Fourier Transform (DFT), this function can be represented by
its $J$ detail scales and its smooth component.
Equation~(\ref{eqn:contwtexpansion}) becomes
\begin{equation}
f(x_i) = c_{0,0} \phi_{0,0}(x_i) + \sum_{j=0}^{J-1} \sum_{l=0}^{2^j-1}
w_{j,l} \psi_{j,l} (x_i),\label{eqn:dwtexpansion}
\end{equation}
and the integrals (\ref{eqn:contwtcoeff}) for the wavelet coefficients
have to be replaced by the appropriate summations.  If the mean of the
function samples $f(x_i)$ is zero, then $c_{0,0}=0$ and the function
can be described entirely in terms of the wavelets $\psi_{j,l}$. As
the scale index $j$ increases from 0 to $J-1$, the wavelets represent
structure of the function on increasingly smaller scales, where each
scale is by a factor of~2 more detailed than the previous one.  The
index~$l$ (which runs from $l=0$ to $2^j-1$) denotes the position of
the wavelet $\psi_{j,l}$ within the $j$th scale level.

If we collect the function samples~$f_i=f(x_i)$ in a column
vector~$\myvec{f}$ (whose length~$N$ must an integer power of~2), then
the DWT (like the DFT) is a linear operation that transforms
$\myvec{f}$ into another vector~$\widetilde{\myvec{f}}$ of the same
length, which contains the wavelet coefficients of the (digitised)
function.  The action of the DWT can therefore be described as a
multiplication of the original vector by the $N\times N$ wavelet
matrix~$\mymatrix{W}$:
\begin{equation}
\widetilde{\myvec{f}} = \mymatrix{W}\ \myvec{f}.
\label{eqn:dwt1d}
\end{equation}
Again like the DFT, the matrix~$\mymatrix{W}$ is orthogonal, and the
inverse transformation can be performed straightforwardly using the
transpose of~$\mymatrix{W}$.  Thus both the DFT and DWT can be
considered as rotations from the original orthonormal basis vectors
$\myvec{e}_i$ in signal space to some new orthonormal basis
$\widetilde{\myvec{e}}_i$ ($i=1,\ldots,N$), with the transformed
vector $\widetilde{\myvec{f}}$ containing the coefficients in this new
basis.

The original basis vectors $\myvec{e}_i$ have unity as the $i$th
element and the remaining elements equal to zero, and hence correspond
to the `pixels' in the original vector~$\myvec{f}$. Therefore
the original basis is the most localised basis possible in real
space. For the DFT, the new basis vectors $\widetilde{\myvec{e}}_i$ are
(digitised) complex exponentials and represent the opposite extreme,
since they are completely non-local in real space but localised in
frequency space. For the DWT, the new basis vectors are the
wavelets, which enjoy the characteristic property of being fairly
localised both in real space and in frequency space, thus occupying an
intermediate position between the original `delta-function' basis
and the Fourier basis of complex exponentials. Indeed, it is
the simultaneous localisation of wavelets in both spaces that makes 
the DWT such a useful tool for analysing data in wide range of applications.

\subsection{The \`{a} trous transform}
\label{sec:wavelet:atrous}

So far we have restricted our attention to discrete wavelet bases that are
orthogonal and non-redundant (i.e. the number of wavelet coefficients
equals the number of points at which the original function is sampled).
By relaxing some of the Daubechies conditions (\ref{eqn:daubcond1})
and (\ref{eqn:daubcond2}), one may represent a sampled
function in terms of wavelet bases that are both non-orthogonal and
redundant. Although this may at first seem a retrograde step, 
\citet*{langer93} have suggested that non-orthogonal, translationally
and rotationally invariant ({\em isotropic}) wavelet transforms are
better suited to the task of image reconstruction than orthogonal
ones. 

The {\em \`a trous} (`with holes') algorithm
(\citealt{holschneider89}; \citealt*{bijaoui94}) is an example of a
non-orthogonal, redundant discrete wavelet transform that is widely
used in image analysis.  Starting from a data vector $c_{J,l}$
($l=1,\ldots,N$, $2^J \le N$) iteratively smoothed vectors are
obtained by
\begin{equation}
c_{j-1,k} = \sum_l H_l c_{j,k+2^{(J-j)} l},
\label{eqn:atrous}
\end{equation}
where each step is effectively a convolution of the image with the
filter mask~$H_l$ using varying step sizes~$2^{J-j}$.  At each scale,
the detail wavelet coefficients contain the difference between the
smoothed image $c_{j-1,k}$ and the image $c_{j,k}$ at the previous
scale:
\[
w_{j-1,k} = c_{j,k} - c_{j-1,k}.
\]
Since no decimation is carried out between consecutive filter steps,
the \`{a} trous transform is redundant. Thus the final wavelet
transformed vector has length~$J \times N$.  The inverse \`{a} trous
transform is simply the sum over the coefficients at all scales:
\[
c_{J,l} = c_{0,l} + \sum_{j=0}^{J-1} w_{j,l}.
\]

\subsubsection{Properties of the \`{a} trous transform}
\label{sec:memwavelet:atrous}

For the \`{a} trous transform the normalisation $\int |\psi(x)|^2\ 
\mathrm{d}x = 1$ no longer holds.  This means that some convenient
properties of the orthogonal transform are lost.  For instance, the
orthogonal wavelet transforms of Gaussian white noise have a constant
dispersion in all wavelet domains.  This is not the case for the \`{a}
trous transform.

As discussed above, the \`{a} trous transform constructs the wavelet
coefficients by successive applications of the same filter mask with
different spacings between pixels, followed by a subtraction of the
smooth component in each step.  This procedure is computationally
efficient because of the compactness of the mask, as the wavelet
coefficients in each domain can be constructed by a sum over only a
small fraction of the coefficients on the previous scale.  Perhaps not
entirely obvious from this construction is that this algorithm is
equivalent to a convolution of the original image with a series of
point spread functions, which do not only have different widths, but
can also assume slightly different shapes on different scales.
Wavelet coefficients are then given by the convolution
\begin{equation}
w_{j-1,k} = \sum_l \psi_{j-1,l-k} c_{J,l}
\label{eqn:atrousconvolution}
\end{equation}
of the input vector and the wavelet function~$\psi_{j-1} (-x)$ at the
$(j-1)$th scale.  For a symmetric wavelet~$\psi_{j-1} (x) = \psi_{j-1}
(-x)$, this is identical to a convolution with the wavelet itself.
Popular choices for the corresponding scaling function~$\phi(x)$ are
the triangle function
\begin{equation}
\phi(x)=\left\{ \begin{array}{cl} 1 - |x| &\mbox{if}\quad x \in
    [-1,1]\\
0 &\mbox{otherwise}
\end{array}\right.
\label{eqn:triangle}
\end{equation}
or a $B_3$-spline \citep*{starck98}.

\begin{figure}
  \begin{center}
    \leavevmode
    \includegraphics[angle=-90,width=8cm]{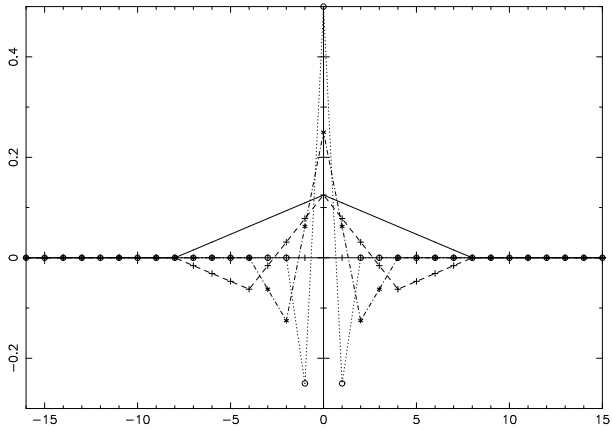}
    \caption{The effective point spread functions of the \`{a} trous
      transform for the triangular scaling
      function~(\ref{eqn:triangle}).  The wavelet coefficients at each
      scale are given by convolutions with the increasingly wider
      wavelet functions (dotted line, dash-dotted line, dashed line).
      The smooth component is given by a convolution with the scaling
      function (solid line).  }
    \label{fig:atrousbasis1}
  \end{center}
\end{figure}
The effective point spread functions for the triangular scaling
function~(\ref{eqn:triangle}) are shown in
Fig.~\ref{fig:atrousbasis1}.  The horizontal axis denotes pixel
offsets.  The first wavelet scale is given by a convolution with the
narrow wavelet function~$\psi(x)$ (dotted line).  The next
scale is given by the same function, but scaled by a factor of~2 in
width and $\frac{1}{2}$ in height (dash-dotted line).  The higher
scales are produced from increasingly wider convolution masks.
Finally, the last scale is the smooth component obtained from a
convolution with the broad scaling function~$\phi(x)$ (solid line).

\begin{figure}
  \begin{center}
    \leavevmode
    \includegraphics[angle=-90,width=8cm]{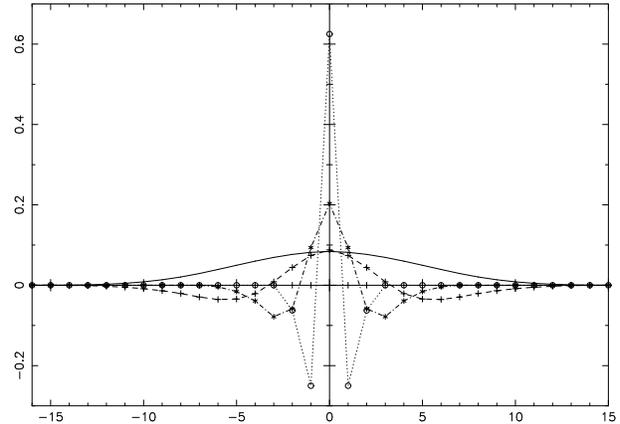}
    \caption{The effective point spread functions of the \`{a} trous
      transform for a $B_3$-spline.  The wavelet coefficients at each
      scale are given by convolutions with the increasingly wider
      wavelet functions (dotted line, dash-dotted line, dashed line).
      The smooth component is given by a convolution with the scaling
      function (solid line).  Compare the triangular mask in
      Fig.~\ref{fig:atrousbasis1}.}
    \label{fig:atrousbasis2}
  \end{center}
\end{figure}
Fig.~\ref{fig:atrousbasis2} shows the effective point spread functions
for the $B_3$-spline given by the convolution mask
\[
\left(\frac{1}{16}, \frac{1}{4},
\frac{3}{8},\frac{1}{4},\frac{1}{16}\right).
\]
Note how the wavelet functions become smoother as their width
increases relative to the pixel resolution.

\subsubsection{The \`{a} trous transform as a Fourier filter}
\label{sec:memwavelet:atrous:filter}

\begin{figure}
  \begin{center}
    \leavevmode
    \includegraphics[angle=-90,width=8cm]{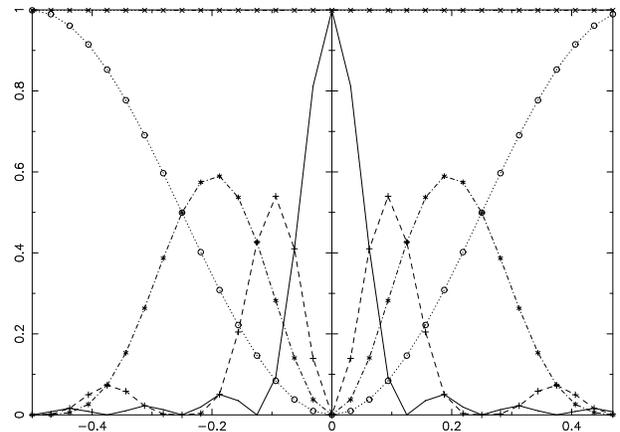}
    \caption{The window functions for the triangular
      transform~(\ref{eqn:triangle}) at different scales in Fourier
      space.  These functions have been obtained by a Fourier
      transform of the functions in Fig.~\ref{fig:atrousbasis1}. The
      horizontal scale denotes spatial frequencies.}
    \label{fig:atrousfft1}
  \end{center}
\end{figure}
The convolution functions shown in Figs.~\ref{fig:atrousbasis1}
and~\ref{fig:atrousbasis2} act as filters corresponding to different
spatial frequency bands.  The corresponding window functions can be
obtained from a Fourier transform of the convolution masks.  The
window functions for the triangular transform from
Fig.~\ref{fig:atrousbasis1} are plotted in Fig.~\ref{fig:atrousfft1}.
The detail coefficients are given by the high-pass filter extending to
the large Fourier modes (dotted line with circles).  The coarser
scales are given by filter functions moving to smaller and smaller
Fourier modes.  The smooth component can be obtained from the low-pass
filter centred around the smallest frequencies (solid line), or
longest wavelengths.  The fact that the inverse transform is simply
given by a sum of the detail scales and the smooth components ensures
that all filters add up to a constant rectangular window of unit height
(crosses on top of the diagram).  The corresponding window functions
for a $B_3$-spline look very similar, but their sidelobes are much
smaller.

\subsection{Two-dimensional wavelet transforms}

The extension of the discrete wavelet transform to two dimensional objects or
images is not unique.  For orthogonal, non-redundant wavelet bases, 
there are two commonly used types of two-dimensional transform.
The terminology used to name these transforms is quite variable, and
different authors even use the same term to refer to different
transforms.  We choose to refer to them as {\em `tensor'} and {\em
  Mallat} (or {\em MRA}) transforms respectively, even though both are
constructed from tensor products of the one-dimensional wavelets.  The
tensor approach simply uses tensor products of the one-dimensional
wavelets as a two-dimensional basis \citep[e.g.][]{press92}.  The MRA
transform was developed by \citet{mallat89} and uses dilated versions
of three different tensor products of one--dimensional wavelets
without mixing scales in different directions.  Both tensor and Mallat
transforms are presented in more detail in
Appendix~\ref{sec:appendix:transform2d}.

The non-orthogonal, redundant \`{a} trous transform lends 
itself particularly well to
generalisations to higher dimensions.  The convolution mask for the
two-dimensional \`{a} trous transform can be obtained from a
two-dimensional rotation of the one-dimensional wavelet function.
One of the most appealing features of the \`{a} trous algorithm is
that it does not single out any special direction in the image plane,
unlike the tensor products that are tailored to vertical, horizontal
and diagonal image features.  In this sense the \`{a} trous transform
is isotropic.  Furthermore, the \`{a} trous transform is also
invariant under translations (the transform of the dilated function is
simply the dilation of the transform). In practice, the convolution
mask for the two-dimensional \`{a} trous transform can even
be obtained from a simple product of two one-dimensional masks,
while retaining the above properties to a sufficient approximation.

\section{Wavelet denoising}
\label{sec:memwavelet:filter}

The application of wavelets to the denoising of CMB maps has recently
been proposed \citep{tenorio99, sanz99a, sanz99b}.  Filtering
techniques exploit the fact that the information content of the
wavelet-transformed image has been compressed into fewer coefficients.
On the other hand, the dispersion of Gaussian white noise is constant
across all wavelet domains after an orthogonal transform, because
white noise has equal power on all image scales.  Thus some wavelet
coefficients become statistically more significant than the original
image pixels, since they are considerably enhanced compared to the
noise.  This observation leads to a filtering scheme where
statistically significant coefficients are kept and the rest are
discarded.  The image obtained by an inverse transform then has much
reduced noise.

A simple realisation of such a filtering scheme uses {\em `hard
  thresholding'}, where each
wavelet coefficient $w_i$ is multiplied by a factor
\begin{equation}
M_i = \left\{ \begin{array}{rl} 1 & \mbox{if}\quad |w_i| \ge \tau_i\\
  0 & \mbox{if}\quad |w_i| < \tau_i 
\end{array} \right. .
\label{eqn:hardthreshold}
\end{equation}
For Gaussian noise, the threshold~$\tau_i$ is usually chosen as some
multiple of the noise dispersion~$\sigma^j_\mathrm{N}$, i.e. $\tau_i =
k \sigma^j_\mathrm{N}$, where the $i$th coefficient lies in the $j$th
wavelet domain.  \citet{starck98} propose to call the
vector~$\myvec{M}$ from~(\ref{eqn:hardthreshold}) the {\em
  multiresolution support}.  A common choice for the threshold is
$k=3$.

For {\em `soft thresholding'}
\begin{equation}
M_i = \left\{ \begin{array}{rl} \frac{w_i - \tau_i}{w_i} & \mbox{if}\quad
      w_i \ge \tau_i \\
  0 & \mbox{if}\quad |w_i| < \tau_i \\
\frac{w_i + \tau_i}{w_i} & \mbox{if} \quad w_i \le \tau_i \end{array} \right.
,
\label{eqn:softthreshold}
\end{equation}
the value~$M_i$ can vary continuously. Coefficients that lie under the
threshold are discarded, whereas the rest are shrunk by the values
$\tau_i$.  Again, there are different prescriptions for choosing
the~$\tau_i$.  One possibility is $\tau_i = k \sigma^j_\mathrm{N}$
where $k$ is of the order of~1. A more advanced method is the
procedure \textit{SureShrink} \citep{donohojohnstone95} which assigns
the threshold~$\tau$ to each resolution level by minimising the Stein
Unbiased Estimate of Risk (SURE) for threshold estimates.  This is
basically a method to minimise the the estimated mean squared error of
the filtered coefficient.  The computational cost of this procedure is
of the order $N \log N$, where $N$ is the number of coefficients in a
domain.  Additionally, one often sets a minimum~$\tau_\mathrm{min}$
and maximum~$\tau_\mathrm{max}$ threshold such that
\textit{SureShrink} is only applied in those wavelet domains~$j$ where
$\tau_\mathrm{min} \le \sigma^j_\mathrm{D}/\sigma^j_\mathrm{N} \le
\tau_\mathrm{max}$.  This avoids damping of the signal for domains of
high signal-to-noise and suppresses noise-dominated coefficients.  The
soft thresholding filter is non-linear, since the values of the filter
coefficients~$M_i$ depend on the data~$w_i$.

The multiresolution support as a mask for regularisation will be
discussed in Section~\ref{sec:memwavelet:alphamodel}.  In
Section~\ref{sec:memwavelet:application}, we will use reconstructions
obtained from wavelet denoising as a benchmark for a comparison with
MEM techniques operating in the wavelet basis.

\section{The maximum-entropy method}
\label{sec:bayes}

The task of recovering the original image from blurred and noisy data
is a typical inverse problem.  This section discusses methods to draw
inferences from data and to solve inverse problems. In particular, we
focus on the maximum-entropy method (MEM) of image reconstruction. We give
here a discussion of the background to the standard MEM technique and then, 
in Section~\ref{sec:memwavelet:transforms}, 
highlight the enhancements to the method provided by performing 
reconstructions in wavelet bases.

\subsection{The inverse problem}
\label{sec:intro:inverse}

In image reconstruction, the inverse problem consists of the task of
estimating the $N_h$ image pixels $h_i$ ($i=1 \ldots N_h$) from the
$N_d$ data samples~$d_i$.  One may find a solution by optimising some
measure of the goodness of fit to the data \citep[see, for example,][]
{titterington85}.  For Gaussian noise, the $\chi^2$-statistic is the
preferred measure:
\begin{equation}
\chi^2(\myvec{h}) = (\mymatrix{R} \myvec{h} - \myvec{d})^\transp
\mymatrix{N}^{-1} (\mymatrix{R} \myvec{h} - \myvec{d}).
\label{eqn:chi2def}
\end{equation}
By minimising~$\chi^2$, the vector~$\hat{\myvec{h}}$ that best fits
the data can be found.

\subsection{Regularisation}

In image reconstructions, the number $N_h$ of parameters is the number
of image pixels, which can be very large.  The parameter estimates are
poorly constrained by the data, and over-fitting leads to a bad
reconstruction of the original image.  In order to avoid over-fitting
and wildly oscillating solutions, some kind of additional information
or regularisation is required.  The regularisation is achieved by an
additional function $S(\myvec{h})$ which penalises `roughness' in the
image. The choice of $S (\myvec{h})$ is determined by what exactly one
considers as roughness.  A compromise between the goodness of fit
$\chi^2(\myvec{h})$ and the regularisation $S(\myvec{h})$ can then be
found by minimising the function
\begin{equation}
F (\myvec{h}) = \tfrac{1}{2} \chi^2(\myvec{h}) - \alpha S(\myvec{h}),
\label{eqn:imagefunc}
\end{equation}
where $\alpha$ is a Lagrange multiplier which determines the degree of
smoothing. As $\alpha$ is varied, solutions lie on a trade-off curve
between optimal fit to the data and maximal smoothness \citep[see, for
example,][]{titterington85}.

\subsection{Bayes' theorem}
\label{sec:intro:bayestheorem}

The previous analysis can be more coherently expressed in a Bayesian
framework. Bayes' theorem can be used as a starting point to draw
statistical inferences from data. It states that the conditional
probability $\Pr (\myvec{h} | \myvec{d})$ for a hypothesis $\myvec{h}$
to be true given some data $\myvec{d}$, the so-called {\em
  `posterior'} probability, is given by
\begin{equation}
\Pr (\myvec{h}|\myvec{d}) = \frac{\Pr(\myvec{d}|\myvec{h})
  \Pr(\myvec{h})}{\Pr (\myvec{d})}. 
\label{eqn:bayestheorem}
\end{equation}
The probability $\Pr(\myvec{d}|\myvec{h})$ is called the {\em
  likelihood} of the data, and $\Pr(\myvec{h})$ is the {\em prior}
probability of the hypothesis.  It can be used to incorporate our
prior beliefs or expectations on possible solutions.  The probability
$\Pr (\myvec{d})$, the {\em evidence}, only depends on the data and
can, for the time being, be viewed as a normalisation constant.

For Gaussian distributed errors on the data points, the likelihood is
then given by
\begin{equation}
\mathcal{L}(\myvec{d} | \myvec{h}) \equiv \Pr (\myvec{d} | \myvec{h}) = \frac{1}{(2\pi)^{N_d/2}
  \sqrt{|\mymatrix{N}|}} e^{- \frac{1}{2} \left( \mymatrix{R} \myvec{h}
  - \myvec{d}\right)^\transp
  \mymatrix{N}^{-1} \left(\mymatrix{R} \myvec{h} - \myvec{d}\right)},
\label{eqn:likelihoodgausserr}
\end{equation}
where $\chi^2 (\myvec{h}) = \left( \mymatrix{R} \myvec{h} -
  \myvec{d}\right)^\transp \mymatrix{N}^{-1} \left( \mymatrix{R}
  \myvec{h} - \myvec{d}\right)$ is the standard misfit statistic
introduced in~(\ref{eqn:chi2def}).  If the parameters are
well-constrained by the data, i.e. the likelihood function is narrow,
the posterior probability will be sufficiently peaked and the errors
on the parameters~$\myvec{h}$ will be small.  Unfortunately, that is
usually not the case in image reconstruction, where the number of
parameters or image pixels is large.  One then has to make use of the
prior $\Pr (\myvec{h})$ to obtain a solution.

\subsection{The maximum entropy method (MEM)}
\label{sec:intro:maxent}

If the likelihood function does not constrain the parameters
sufficiently, the choice of a prior becomes important.  There are many
possible choices for the prior.  The prior is usually of the
exponential form \citep[e.g.][]{skilling89}
\begin{equation}
\Pr (\myvec{h}) \propto \exp \left[ {\alpha S(h)} \right],
\label{eqn:prior}
\end{equation}
where $S(\myvec{h})$ is a regularisation function and $\alpha$ is a
constant.  Assuming a likelihood function given
by~(\ref{eqn:likelihoodgausserr}), the posterior probability
from~(\ref{eqn:bayestheorem}) then becomes
\begin{equation}
\Pr (\myvec{h}|\myvec{d}) \propto \exp \left[{-\frac{1}{2} \chi^2(\myvec{h}) +
  \alpha S(\myvec{h})} \right].
\label{eqn:posterior}
\end{equation}
Clearly, the optimal choice of the regularisation has to reflect our
knowledge of the expected solution \citep[see][]{frieden83}.  A list
of commonly used regularisation functions can be found in
\citet{titterington85}.  These are either functions that are quadratic
in the hypothesis $\myvec{h}$ or that use some kind of logarithmic
entropy.

An efficient prior is provided by the (Shannon) information entropy of
the image.  Restricting ourselves to images whose pixel values are
strictly positive, the image can be considered as a positive additive
distribution (PAD).  The entropy function~(\ref{eqn:gullskilling}) is
a generalisation of the Shannon entropy.  The measure $\myvec{m}$ is
often called the {\em `model'}, because the entropy is maximised by
the default solution $h_i=m_i$ ($i=1 \ldots N_h$).  Given an entropy
function~$S(\myvec{h})$, the posterior probability can be maximised by
minimising its negative logarithm
\begin{equation}
- \ln [\Pr (\myvec{h} | \myvec{d}) ] = F(\myvec{h}) = \frac{1}{2}
  \chi^2(\myvec{h}) - \alpha S(\myvec{h}). 
\label{eqn:logposterior}
\end{equation}
This is the maximum entropy method
\citep[MEM; see, for example,][]{gull89,skilling89,gullskilling99}.

In~(\ref{eqn:logposterior}), we have omitted a constant additive
term that comes from the normalisation of the posterior probability.
This term includes the logarithm of the evidence $\Pr(\myvec{d})$.
The evidence becomes useful because it is conditional on the
underlying assumptions, for example the values of the
constant~$\alpha$ and the model~$\myvec{m}$, and can be written as
$\Pr (\myvec{d} | \alpha, \myvec{m},\ldots)$.  From Bayes'
theorem~(\ref{eqn:bayestheorem}), the posterior
probability~$\Pr(\alpha, \myvec{m},\ldots | \myvec{d})$ for the
model~$\myvec{m}$ and regularisation~$\alpha$ depends on the evidence
$\Pr(\myvec{d}| \alpha, \myvec{m}, \ldots)$:
\begin{equation}
\Pr (\alpha, \myvec{m},\ldots | \myvec{d}) = \frac{\Pr(\myvec{d}|
  \alpha, \myvec{m},\ldots) \Pr(\alpha, \myvec{m},\ldots)}
{\Pr(\myvec{d})}.
\label{eqn:evidence}
\end{equation}
Thus, in the same way as the likelihood discriminates between
hypotheses the evidence helps to discriminate between different priors
or models \citep[again, e.g.][]{gullskilling99}.  In
Section~\ref{sec:intro:maxent:alpha}, the evidence will be used to set a
Bayesian value of the regularisation constant~$\alpha$.  In image
reconstructions the model $\myvec{m}$ is often set uniformly across
the image, to the value of the expected image background.  Again, for
a more refined analysis the evidence can be used to discriminate
between models.

\subsection{The regularisation parameter $\alpha$}
\label{sec:intro:maxent:alpha}

The parameter $\alpha$ introduced in~(\ref{eqn:prior}) determines the
amount of regularisation on the image.  It is clear that minimising
$\chi^2$ only (by setting $\alpha=0$) would lead to a closer agreement
with the data, and thus to noise-fitting. On the other hand,
maximising the entropy alone by setting $\alpha=\infty$ would lead to
an image which equals the default everywhere.  Indeed, for every
choice of $\alpha$ there is an image $\myvec{h}(\alpha)$ corresponding
to the minimum of~$F(\myvec{h})$ for that particular choice. The
images~$\myvec{h}(\alpha)$ vary along a trade-off curve as $\alpha$ is
varied.

There are several methods for assigning an optimal value to $\alpha$.
In early MEM applications, $\alpha$ was chosen such that for the final
reconstruction~$\hat{\myvec{h}}$ the misfit statistic $\chi^2$
equalled its expectation value~$\chi^2 (\hat{\myvec{h}}) = N_d$, i.e.
the number~$N_d$ of data values.  This choice is often referred to as
{\em historic MEM}. It can be shown that it leads to systematic
underfitting of the data \citep{titterington85}. However, an optimum
value of $\alpha$ can be assigned within the Bayesian framework itself
\citep{gull89,gullskilling99}.  Treating $\alpha$ as another parameter
in the hypothesis space, or rather as a part of the model or theory
that the parameter~$\myvec{h}$ belongs to, one can remove the dependence of
the posterior on this parameter by marginalising over~$\alpha$:
\[
\Pr (\myvec{h} | \myvec{d}) = \int \Pr(\myvec{h}| \myvec{d}, \alpha)
\Pr(\alpha|\myvec{d})\ \mathrm{d}\alpha.
\]
From~(\ref{eqn:evidence}), we have $ \Pr(\alpha|\myvec{d}) \propto
\Pr(\myvec{d}| \alpha) \Pr(\alpha)$.  If the evidence~$\Pr (\myvec{d}
| \alpha)$ is sufficiently peaked, it will overwhelm any priors
on~$\alpha$ and one can simply use the optimal value~$\hat{\alpha}$
which maximises the evidence.  This choice of $\alpha$ is called {\em
  classic maximum entropy} \citep{gull89}. It can be shown that the
Bayesian value of $\alpha$ may be reasonably approximated by choosing
its value such that the value of $F (\myvec{h})$ at its minimum is
equal to half the number of data points, i.e. $F \approx N_d /2$
\citep{mackay92a}.

\subsection{Errors on the maximum entropy estimates}
\label{sec:intro:maxent:errors}

There are two principal methods of quantifying errors on maximum
entropy reconstructions.  Firstly, one may evaluate the Hessian
matrix $\mymatrix{H}= \nabla_\myvec{h} \nabla_\myvec{h} F$ at the
optimal $\hat{\myvec{h}}$, from which the covariance matrix of the
parameters is given by $\mymatrix{C} = \mymatrix{H}^{-1}$.  However,
this typically requires the inversion of a large matrix, which is
unfortunately non-diagonal and so requires a large computational
effort.  Secondly, one may take `samples' from the posterior
probability distribution and quantify the error on any particular
parameter $h_i$ (or combinations of parameters) by examining how $h_i$
varies between samples from the distribution \citep{gullskilling99}.
Additionally, when testing the method on simulations, it is always
possible to quantify the errors on the reconstruction by using a
Monte-Carlo approach. Since this is in fact the most robust method, we
adopt it in Section~\ref{sec:memwavelet:application}, in which we
analyse some simulated data.

\subsection{The intrinsic correlation function (ICF)}
\label{sec:intro:maxent:icf}

One of the fundamental axioms on which maximum entropy is based is
that it should not by itself introduce correlations in the
reconstructions \citep{gullskilling99}.  However, it is often the case
that a priori there is reason to believe that for a specific
application the reconstructed quantities are not uncorrelated.
Fortunately, any additional information on the correlation structure
can be exploited to improve the quality of the reconstruction.

\begin{figure}
  \begin{center}
    \leavevmode
    \includegraphics[width=7cm]{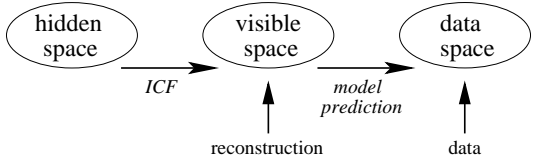}
    \caption{The relationship between data, visible and hidden space.}
    \label{fig:icf}
  \end{center}
\end{figure}
Denoting the reconstructed image by~$\myvec{v}$ instead
of~$\myvec{h}$, the correlation structure of $\myvec{v}$ can be
encoded in a function~$\mymatrix{K}$ called the intrinsic correlation
function (ICF).  Using the ICF~$\mymatrix{K}$, the reconstructed ({\em
`visible'}) image~$\myvec{v}$ can be derived from a set of new ({\em
`hidden'}) parameters $\myvec{h}$ by
\begin{equation}
\myvec{v} = \mymatrix{K} \myvec{h}.
\label{eqn:icf}
\end{equation}
Instead of reconstructing the image vector directly, the entropy is
maximised with respect to the {\em `hidden image'}~$\myvec{h}$.  The
elements of~$\myvec{h}$ are ideally a priori uncorrelated and of unit
variance; the information on the correlations is absorbed
into~$\mymatrix{K}$.  The relationship between data, visible and
hidden image is depicted in Fig.~\ref{fig:icf}.  The visible
reconstruction can be derived from the hidden image through the ICF,
and the predicted noiseless data can be calculated using the response
matrix and is given by $\myvec{d} = \mymatrix{R}\myvec{v}$.

If the covariance matrix $\mymatrix{C}= \langle v_i v^\dagger_j \rangle$ is known, the ICF can be constructed
straightforwardly by a diagonalisation or Cholesky decomposition
of~$\mymatrix{C}$.  
Unfortunately the correlation structure is usually not known in advance, and a suitable ICF
has to be found from empirical or heuristic criteria.

\section{Transforming the inverse problem}
\label{sec:memwavelet:transforms}

Without knowledge of the correlation structure, the ICF will have to
be based on some assumptions about the correlations.  One may prefer
to find an ICF that works specifically well in certain cases, or one
that performs well in the worst case.  In the following, we will
investigate several types of wavelet transforms as ICFs.
  
Let us consider more closely the relationship between the data, visible
and hidden image vectors. The $N_\mathcal{D}$-dimensional data
vector~$\myvec{d}$ is an element of the data space~$\mathcal{D}$.
Similarly, the visible vector $\myvec{v} \in \mathcal{V}$, $\dim
\mathcal{V} = N_\mathcal{V}$ and the hidden vector $\myvec{h} \in
\mathcal{H}$, $\dim \mathcal{H} = N_\mathcal{H}$.  The data vector is
then given by the transform
\begin{equation}
\begin{array}{ccccc}
\myvec{h} &\longmapsto &\myvec{v}&
\longmapsto &\myvec{d}\\
\in&&\in&&\in\\
\mathcal{H} &\stackrel{{\mymatrix{K}}}{\longrightarrow} &\mathcal{V}&
\stackrel{\mymatrix{R}}{\longrightarrow}& \mathcal{D},
\end{array}
\label{eqn:memtransforms}
\end{equation}
where the ICF~$\mymatrix{K}$ is a wavelet transform.  In practice, we
will use the transpose~$\mymatrix{K} = \mymatrix{W}^\transp$ of the
wavelet transform in this step, i.e. $\myvec{v} = \mymatrix{W}^\transp
\myvec{h}$.  For orthogonal wavelet transforms, $\mymatrix{W}^\transp
= \mymatrix{W}^{-1}$ and this choice of~$\mymatrix{K}$ ensures that
the transformation $\myvec{v} \mapsto \myvec{h}$ is given by the
wavelet transform and the hidden space simply consists of the wavelet
coefficients of the reconstruction.  For non-orthogonal transforms,
however, $\mymatrix{W}^\transp = \mymatrix{W}^{-1}$ does not hold.
  The $\chi^2
(\myvec{v})$-function is defined on the space $\mathcal{V}$ of visible
vectors $\myvec{v}$ and the entropy function $S(\myvec{h})$ on
$\mathcal{H}$.  There are now two ways to construct a MEM
algorithm.
\begin{enumerate}
\item The first method is to choose an entropy function $S(\myvec{h})$
  in the space $\mathcal{H}$ of hidden images and to maximise the
  functional $F(\myvec{h}) = \frac{1}{2}\chi^2 (\mymatrix{K}\myvec{h})
  - \alpha S(\myvec{h})$.  In the following, we will call this
  approach {\em `wavelet MEM'}. Fig.~\ref{fig:wavemem} shows a
  schematic depiction of wavelet MEM.  It is the straightforward way
  to implement the MEM in a hidden-space algorithm.  A hidden-space
  MEM kernel is, for example, given in the software package
  \textsc{memsys5} \citep{gullskilling99}.  We also note that a
  numerical implementation requires the evaluation of derivatives
  of~$F$.  These are discussed in Appendix C and, 
from~(\ref{eqn:chi2gengrad}), we see that the transpose of
  $\mymatrix{K}$ is required for the evaluation of the gradient.  This
  is why $\mymatrix{K} = \mymatrix{W}^\transp$ is a better choice than
  $\mymatrix{K} = \mymatrix{W}^{-1}$ for non-orthogonal transforms.
  Furthermore, in Section~\ref{sec:memwavelet:icfatrous} we will show
  that the use of the transpose has a very simple interpretation for
  the non-orthogonal \`{a} trous transform.
\begin{figure}
  \begin{center}
    \leavevmode
    \includegraphics[width=7cm]{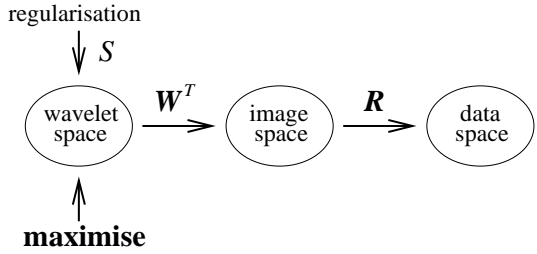}
    \caption{Schematic depiction of wavelet MEM. The data are
      predicted by a composition of the transpose of the wavelet
      transform~$\mymatrix{W}^\transp$ and the instrumental
      response~$\mymatrix{R}$.  The entropic regularisation is applied
      to the hidden (wavelet) space, and the posterior functional is
      maximised with respect to the wavelet coefficients.}
    \label{fig:wavemem}
  \end{center}
\end{figure}
\item Another alternative is to define $F$ in terms of the visible
  space quantities: $F(\myvec{v}) = \frac{1}{2} \chi^2 (\myvec{v}) -
  \alpha S(\mymatrix{K}^\transp \myvec{v})$. (Since we defined
  $\mymatrix{K} = \mymatrix{W}^\transp$, the transform applied to the
  vector~$\myvec{v}$ in the entropy term is simply the wavelet
  transform.)  This method has been pursued by, for instance,
  \citet{pantinstarck96} and \citet{starck01}.  In fact, the entropy
  $S(\mymatrix{W} \myvec{v})$ can be viewed as a function on
  $\mathcal{V}$, and its proponents call it the `multiresolution
  entropy'. We will call this approach {\em `wavelet regularised MEM'}.
  An illustration is shown in Fig.~\ref{fig:waveregmem}.    The
  transform~$\mymatrix{K}$ is not an ICF in the same sense as
  discussed in Section~\ref{sec:intro:maxent:icf}, since the parameters
  that are reconstructed by the MEM, the visible image pixels, are
  still correlated.  We also note that for redundant wavelet
  transforms, this approach has the advantage that the numerical
  minimisation problem is lower dimensional, since $N_\mathcal{V} <
  N_\mathcal{H}$.
\begin{figure}
  \begin{center}
    \leavevmode
    \includegraphics[width=7cm]{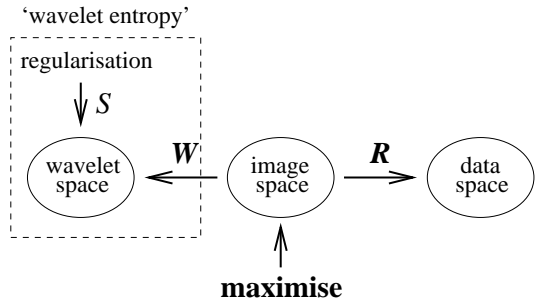}
    \caption{Schematic depiction of wavelet regularised MEM.   The
      entropic regularisation is applied to the wavelet coefficients
      obtained from the image, but the posterior functional is
      maximised with respect to the image pixels.  The wavelet
      transform and entropy functional act as a new effective `wavelet
      entropy'.}
    \label{fig:waveregmem}
  \end{center}
\end{figure}
\end{enumerate}

Despite the use of the same linear transform~$\mymatrix{K}$, `wavelet
regularised MEM' is not entirely equivalent to `wavelet MEM' because
of the non-linearity of the entropy functional.  In
Appendix~\ref{sec:wavemem:equivalence}, we show that both methods become
equivalent for orthogonal wavelet transforms if a quadratic
regularisation function is used.

The type of wavelet transform is not the only free parameter one has
to pick in wavelet MEM.  For instance, there is still freedom in the
choice of the entropy functional~$S$.  As mentioned before, an optimal
choice should reflect the expected distribution of the hidden image
vectors~$\myvec{h}$.  A more detailed discussion
of entropy functionals used with wavelets and maximum entropy is given
in Section~\ref{sec:memwavelet:entropy}. 

There is no reason in the MEM that the data space~$\mathcal{D}$ and
visible space~$\mathcal{V}$ be identical.  For example,
\citet{bridle98} and \citet{marshall02} have applied maximum entropy
to the reconstruction of lensing mass profiles of galaxy clusters from
shear or magnification data.  In this case, $\mathcal{D}$ and
$\mathcal{V}$ do not even share the same physical dimensions.
However, in image reconstruction the data image and the reconstruction
are often not only given in the same physical units, but on the same
discretised grid, and we have $\mathcal{D} = \mathcal{V}$.  In this
case, it is possible to incorporate information gleaned directly from
the data into the choice of the regularisation.  The ICF~$\mymatrix{K}
(\myvec{d})$ and the entropy $S(\myvec{h}, \myvec{d})$ may become
explicitly data-dependent.  In practice, the data dependence is
usually introduced by a modification of the model~$\myvec{m}$.

\subsection{The entropy function}
\label{sec:memwavelet:entropy}

From~(\ref{eqn:gullskilling}), for a positive additive
distribution~$\myvec{h}$, the cross-entropy $S_+(\myvec{h},\myvec{m})$
of the image~$\myvec{h}$ with some model~$\myvec{m}$ of the image is
given by the sum
\[
S_+(\myvec{h}, \myvec{m}) = \sum_i s_+ (h_i, m_i),
\]
where
\begin{equation}
s_+(h, m) = h - m - h \log \frac{h}{m}.
\label{eqn:posentropy}
\end{equation}
The function~$s_{+}(h,m)$ reaches a global maximum of zero at $h=m$.
Thus, in the absence of data, the reconstruction takes on the default
value~$m$ for all pixels; in practice, the value of~$m$ is often set
to the level of the expected image background.

Many astronomical images, such as maps of CMB fluctuations, generally
consist of both positive and negative pixels.  Furthermore, even if an
image were purely positive, some of its wavelet coefficients may be
negative. The entropy~(\ref{eqn:posentropy}) is thus inapplicable to
such images.  A simple generalisation of~(\ref{eqn:posentropy}) to
negative values is
\begin{equation}
s_{| \cdot |}(h,m) = |h| - m - |h| \log\frac{|h|}{m},
\label{eqn:psentropy}
\end{equation}
which has been used by \citet{pantinstarck96}.  This function does not
have a unique maximum, since both $h=\pm |m|$ maximise the entropy.
For $h\rightarrow 0$, one finds $s_{| \cdot |}(h) \rightarrow -m$.
However, the entropy is not defined at zero, which is generally the
expected default state of the image if the mean has been subtracted.
In practice, the model values~$m$ have to be close to zero and small
compared to any realistic data in order to avoid the introduction of a
spurious background signal.  \citet{pantinstarck96} use the value $m =
k \sigma$, where $\sigma$ is the \rms\ of the data and $k =
\frac{1}{100}$ is an arbitrarily chosen, small constant.

The proper way to extend the entropy~(\ref{eqn:posentropy}) to images
that take both positive and negative values is the entropy definition
\citep{hobsonlasenby98,gullskilling99}
\begin{equation}
s_{\pm} (h,m) = \Psi - 2 m - h \log \frac{\Psi + h}{2 m},
\label{eqn:posnegentropy}
\end{equation}
where $\Psi = \sqrt{{h}^2 + 4 {m}^2}$.  The entropy~$s_{\pm}$ has a
maximum at~$h=0$.  In the positive/negative
entropy~(\ref{eqn:posnegentropy}) the role of the model~$m$ is
different from that in~(\ref{eqn:posentropy}).  The value of~$m$
determines the width of the entropy function and thus controls the
magnitude of the allowed deviations from the default value.  Hence the
model can be considered as a level of `damping' imposed on the image.
From a dimensional analysis, the obvious choice for~$m$ is to set it
to the expected signal \rms.  If data and visible space are identical
and the signal-to-noise ratio is sufficiently high, the \rms\ of the
observed data can be a good approximation to that of the signal.

\begin{figure}
  \begin{center}
    \includegraphics[angle=-90,width=8cm]{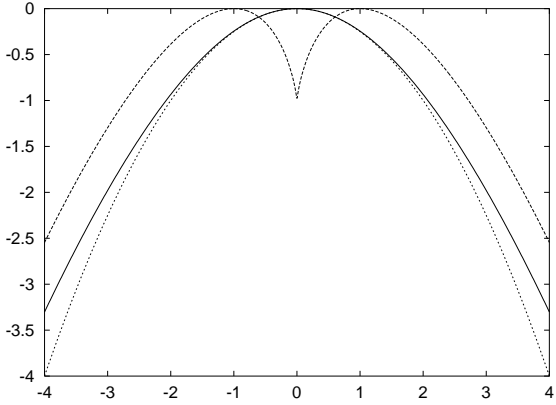}
    \caption{The entropy function $s_{\pm} (h)=\psi - 2m-h
      \log\frac{\psi+h}{2m}$ (solid line), its quadratic approximation
      $s(h) = -h^2/(4m)$ (short dashed line) and $s_{| \cdot |}(h) =
      |h|-m-|h| \log\frac{|h|}{m}$ (long dashed line). In all cases,
      the model was chosen to be $m=1$. The function~$s_{| \cdot |}$
      is not defined at $h= 0$. }
    \label{fig:entropy}
  \end{center}
\end{figure}
Fig.~\ref{fig:entropy} shows the entropy functions $s_{| \cdot |}$ and
$s_{\pm}$.  From the definitions~(\ref{eqn:psentropy})
and~(\ref{eqn:posnegentropy}), we see that, for a given model, both
functions differ only by a constant offset in the limit of large~$h$:
$s_{| \cdot |}(h) \rightarrow s_{\pm} (h)+m\ (h \rightarrow \infty)$.
However, in order to minimise the cusp of $s_{| \cdot |}$ around zero,
the models will generally be chosen differently, and the maximum of
$s_{| \cdot |}$ will be significantly narrower than that of $s_{\pm}$.

Expanding $s_{\pm}$ around zero
\begin{eqnarray}
s_\pm(h,m) & = & \sum_{n=1}^\infty \frac{(\prod_{i=1}^n
  (2i-3))^2 (-1)^n}{(2m)^{2n-1}} \frac{h^{2n}}{(2n)!} \nonumber\\ 
& = & - \frac{h^2}{2 (2m)} +  \frac{h^4}{4! (2m)^3} - \frac{3^2 h^6}{6!
  (2m)^5}
+ \frac{(3 \cdot 5)^2 h^8}{8! (2m)^7} - \ldots\nonumber\\
& =& - \frac{h^2}{4m} + \mathcal{O} (h^4),
\label{eqn:expandentropy}
\end{eqnarray}
one obtains the quadratic approximation
\begin{equation}
s_2 (h) = - \frac{h^2}{4 m},
\label{eqn:quadentropy}
\end{equation}
which is plotted in Fig.~\ref{fig:entropy}.  With this entropy, MEM
reduces to a scaled least squares.  The quadratic entropy can also be
obtained from the large-$\alpha$ limit $h \simeq m$ of the standard
entropy~$s_+(h)$, but with a different scale factor~$\frac{1}{2m}$.
In a MEM reconstruction, one evaluates the product $\alpha S$ of the
entropy~$S$ and a regularisation constant~$\alpha$.  The
constant~$\alpha$ is not dimensionless; its dimension~$[\alpha] =
1/[h]$ is given by the dimension~$[h]$ of~$h$.  If the model is chosen
proportional to the signal \rms~$\sigma_\mathrm{S}$, then,
from~(\ref{eqn:expandentropy}), the product $\alpha s$ becomes
invariant under a rescaling of~$h$ if $\alpha$ is also rescaled by
$1/\sigma_\mathrm{S}$.  In fact, to first order, any change in the
model~$m$ can be absorbed by a reciprocal change in the regularisation
constant~$\alpha$:
\begin{equation}
\alpha s \propto - \frac{\alpha}{m} {h^2}. 
\label{eqn:equivalphamodel}
\end{equation}
This explains why $s_{| \cdot |}$ produces reconstructions similar to
those obtained for to~$s_{\pm}$ despite of its narrow shape for small
models.  We note that in a more recent work, \citet{starck01} use the quadratic
approximation~$s_2$.

Finally we note that \citet{pantinstarck96} propose to choose the
regularisation parameter~$\alpha$ dependent on the scale or even the
image pixel, instead of setting a constant $\alpha$ globally.  This
can be achieved by introducing an additional weighting factor
${\alpha}_i$ for each pixel $i$:
\begin{equation}
S (\myvec{h},\myvec{m}) = \sum_i {\alpha}_i \ s(h_i, m_i).
\label{eqn:posnegalphaentropy}
\end{equation}
However, from~(\ref{eqn:equivalphamodel}) we see that this
modification is again to first order equivalent to a corresponding
change in the model value~$m_i$, and so can be achieved by adopting a
non-constant model.

\subsection{Choosing the parameters $\alpha$ and $m$}
\label{sec:memwavelet:alphamodel}

The choice of the regularisation parameter~$\alpha$ has been discussed
in Section~\ref{sec:intro:maxent:alpha}.  Classic maximum entropy uses
a Bayesian choice of~$\alpha$, as implemented in the software
package~\textsc{memsys5}.  For wavelet regularised MEM, the
maximisation takes place in visible space and the Hessian of the
effective entropy is not diagonal, which means it cannot be
implemented in \textsc{memsys5}.  As mentioned above,
\citet{pantinstarck96} propose to choose a pixel-dependent value
of~$\alpha$ in order to enhance or alleviate the regularisation on
certain coefficients. They use
\begin{equation}
{\alpha}_i \propto {\sigma^j_\mathrm{N}}(1 - M_i) 
\label{eqn:psalpha}
\end{equation}
in~(\ref{eqn:posnegalphaentropy}), where $M_i$ is the multiresolution
support~(\ref{eqn:hardthreshold}) defined in
Section~\ref{sec:memwavelet:filter}, the $i$th pixel is assumed to lie
in the $j$th~wavelet domain and the factor~$\sigma^j_\mathrm{N}$ is
the noise dispersion in the $j$th~wavelet domain.  The factor
$(1-M_i)$ reduces the regularisation of coefficients with a good
signal-to-noise, and the factor~$\sigma^j_\mathrm{N}$ introduces an
additional pixel-dependent regularisation.

The pixel-dependent factor~${\alpha}_i$ is scaled by the global
parameter~$\alpha$.  In the following, we will simply employ the
historic maximum entropy criterion~$\chi^2=N_\mathcal{D}$ to fix a
global~$\alpha$ in cases where \textsc{memsys5} cannot be used to
determine a Bayesian value.

In real-space MEM, there is often no a priori reason to assign a
varying model to particular image regions, since in the absence of any
additional information one would expect a constant signal dispersion
across the image.  The wavelet transform, however, is designed to
enhance the signal-to-noise ratio on some wavelet coefficients to
allow a sparse representation of the signal.  Consequently one would
expect different signal dispersions in each wavelet domain, and a
uniform choice of the model seems appropriate only within domains.  If
data and image space are identical, the signal dispersion in the
$j$th~wavelet domain~$\sigma^j_\mathrm{S}$ can be estimated from the
dispersion of the data~$\sigma^j_\mathrm{D}$ and of the noise
$\sigma^j_\mathrm{N}$:
\[
\sigma^j_\mathrm{S} = \sqrt{\sigma^{j 2}_\mathrm{D}-\sigma^{j 2}_\mathrm{N}}.
\]
Now the model can be set to $m_i = \sigma^j_\mathrm{S}$, where the
$i$th pixel lies in the $j$th wavelet domain.  If no analytic noise
model is available, the noise dispersion can be obtained from
Monte-Carlo simulations.  

Other models have been proposed that involve the use of
signal-to-noise ratios rather than signal dispersions.  For orthogonal
transforms, the dispersion of the wavelet coefficients of white noise
is scale-invariant and the signal-to-noise is proportional to the
signal dispersions.

\subsection{Choice of the wavelet basis}
\label{sec:memwavelet:basis}

For a given data set, one has to chose a specific wavelet basis for a
MEM reconstruction.  Some bases will obviously be more suitable to fit
the data than others, and one naturally wants to find a way to choose
the best one.  A common selection criterion is the information entropy
of the image in the wavelet representation
\citep{coifmanwickerhauser92,zhuangbaras94,hobson99}.  As will be
discussed in Section~\ref{sec:memwavelet:application:einstein}, we find
empirically that for orthogonal wavelets, the choice of the wavelet
basis in the MEM algorithm does not play an important role.  Only Haar
wavelets seem to perform significantly worse than other types.

\section{Multi-resolution reconstruction}
\label{sec:memwavelet:multichannel}

The advantage of transforming the inverse problem to wavelet space
is that it allows for a natural multiresolution
description of the image. Indeed, our approach may be
be considered in the context of the multi-channel ICF 
proposed by \citet{weir92}, which we now discuss.

\subsection{Multi-channel ICF}

Traditionally the ICF ~$\mymatrix{K}$ is taken to be a convolution
with some point spread function~$P(x)$.  Thus the visible image is a
blurred version of the hidden variables.  Popular point spread
functions comprise B-splines of various orders \citep{gullskilling99}
or Gaussians \citep[e.g.][]{weir92}.  One deficiency of the `blurring'
ICF is that the width of the point spread function introduces a
characteristic scale for the pixel correlations.  In most
applications, however, objects and correlations of varying sizes and
scales are present in the same image.  A straightforward
generalisation of the blurring ICF is the multi-channel ICF
\citep{weir92}.
\begin{figure}
  \begin{center}
    \leavevmode
    \includegraphics[width=8cm]{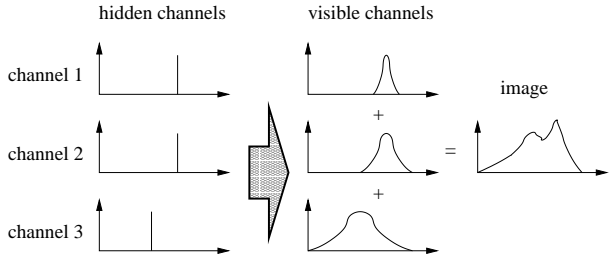}
    \caption{The multi-channel ICF.  The hidden image consists of
      several images or channels (left).  The visible channels are
      obtained by convolutions of the hidden parameters with point
      spread functions of different widths (middle).  The visible
      image is a weighted sum of the image channels (right).  }
    \label{fig:multichannelicf}
  \end{center}
\end{figure}
This method is illustrated in Fig.~\ref{fig:multichannelicf}.  The
hidden variables consist of a number of different images or
channels~$\myvec{h}_i$.  Each hidden channel is blurred with a
different point spread function~$\mymatrix{K}_i$ such that
$\myvec{v}_i = \mymatrix{K}_i \myvec{h}_i$.  The visible
image~$\myvec{v}$ is obtained by a weighted sum $\myvec{v} = \sum_i
w_i \myvec{v}_i$ of all blurred image channels $\myvec{v}_i$, where
the weight on the $k$th channel is denoted by $w_i$.  The entropy
function is applied to the hidden variables.  If the weights on
different scales are chosen appropriately, it is entropically
favourable to represent extended structure in the visible image by a
single or very few coefficient in the hidden domain.

One weakness of this approach is that it introduces a large number of
new free parameters, like the widths of the point spread functions and
the weighting factors~$w_k$ of the different channels (and a complete
new hidden image for each scale).  If there is a priori knowledge on
the expected correlation scales of objects, the width of the point
spread function can be chosen accordingly.  However, this will rarely
be the case.  The reconstructions can become strongly sensitive to the
chosen set of ICF widths, the number of channels, the weights and the
models in different channels \citep{bontekoe94}.  In the `pyramidal
MEM' \citep{bontekoe94}, the different channels contain different
numbers of hidden parameters.  With decreasing resolution, the number
of parameters in each consecutive channel is reduced by a factor of a
half in each image dimension.  \citet{bontekoe94} find empirically
that uniform models and weights (i.e. $w_k = 1$ for all channels~$k$)
and the same point spread function (scaled by factors of~2) can be
used for all channels.  They also note that the pyramid images can be
interpreted as spatial band-pass filters.

We note out that the weights~$w_k$ of the visible image channels play
a similar role as the scale-dependent regularisation
parameters~${\alpha}_i$ from Section~\ref{sec:memwavelet:entropy} that
were proposed by \citet{pantinstarck96}.  The different weights
correspond to a rescaling of the hidden image channels.
From~(\ref{eqn:equivalphamodel}), this is again to first order
equivalent to a corresponding rescaling (albeit quadratically) of the
regularisation~$\alpha$ or the model~$m$.

\subsection{The \`{a} trous transform as ICF}
\label{sec:memwavelet:icfatrous}

If one writes the non-orthogonal \`{a} trous transform as a
series of convolutions with scaling or wavelet functions, as discussed
in Section~\ref{sec:wavelet:atrous}, one can show that, for the \`{a}
trous transform, the ICF~$\mymatrix{K} = \mymatrix{W}^\transp$ used in
wavelet MEM is just a special case of a multi-channel ICF.

\begin{figure}
  \begin{center}
    \leavevmode
    \begin{tabular}{ll}
      (a)&\includegraphics[width=6cm]{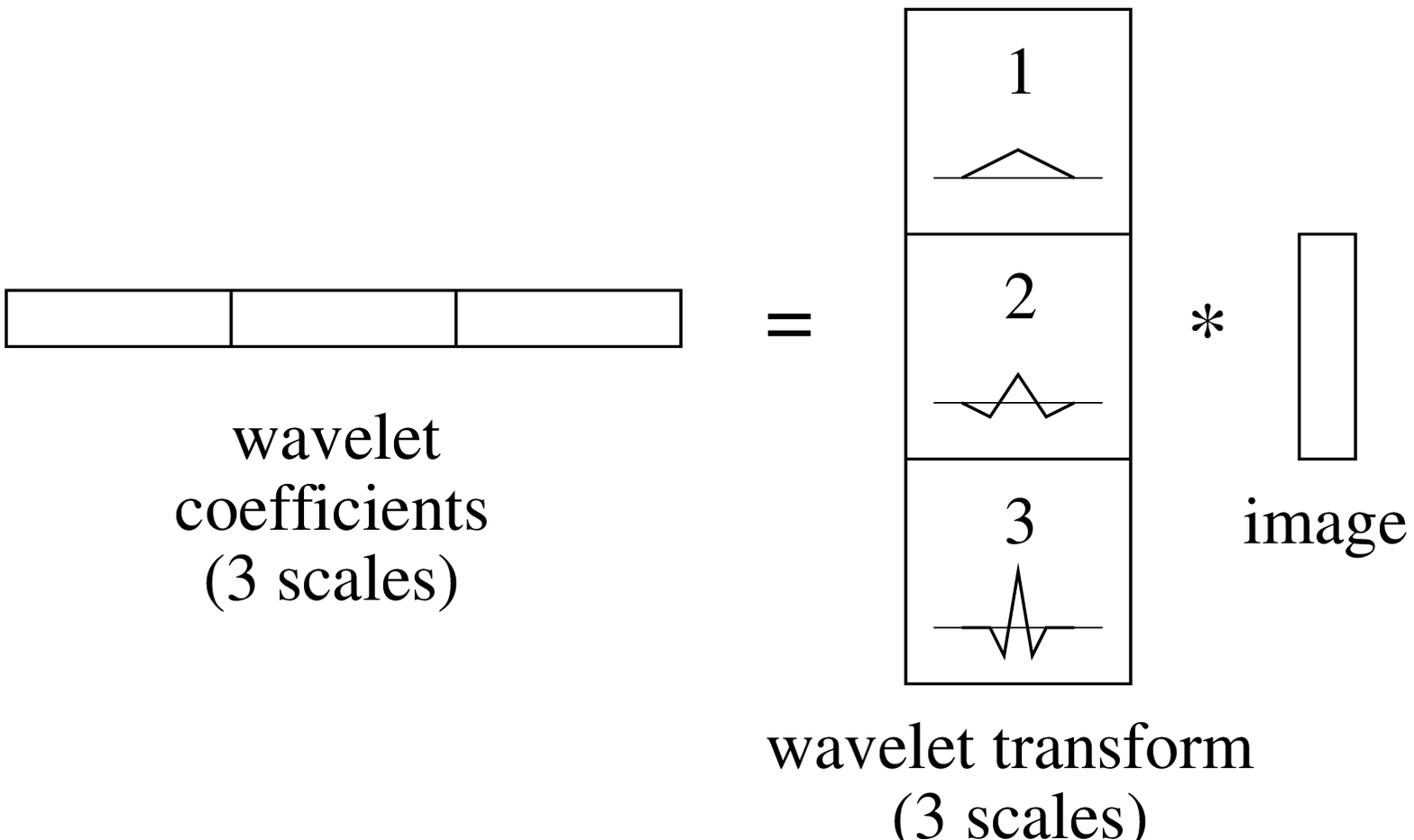}\\
      (b)&\includegraphics[width=6cm]{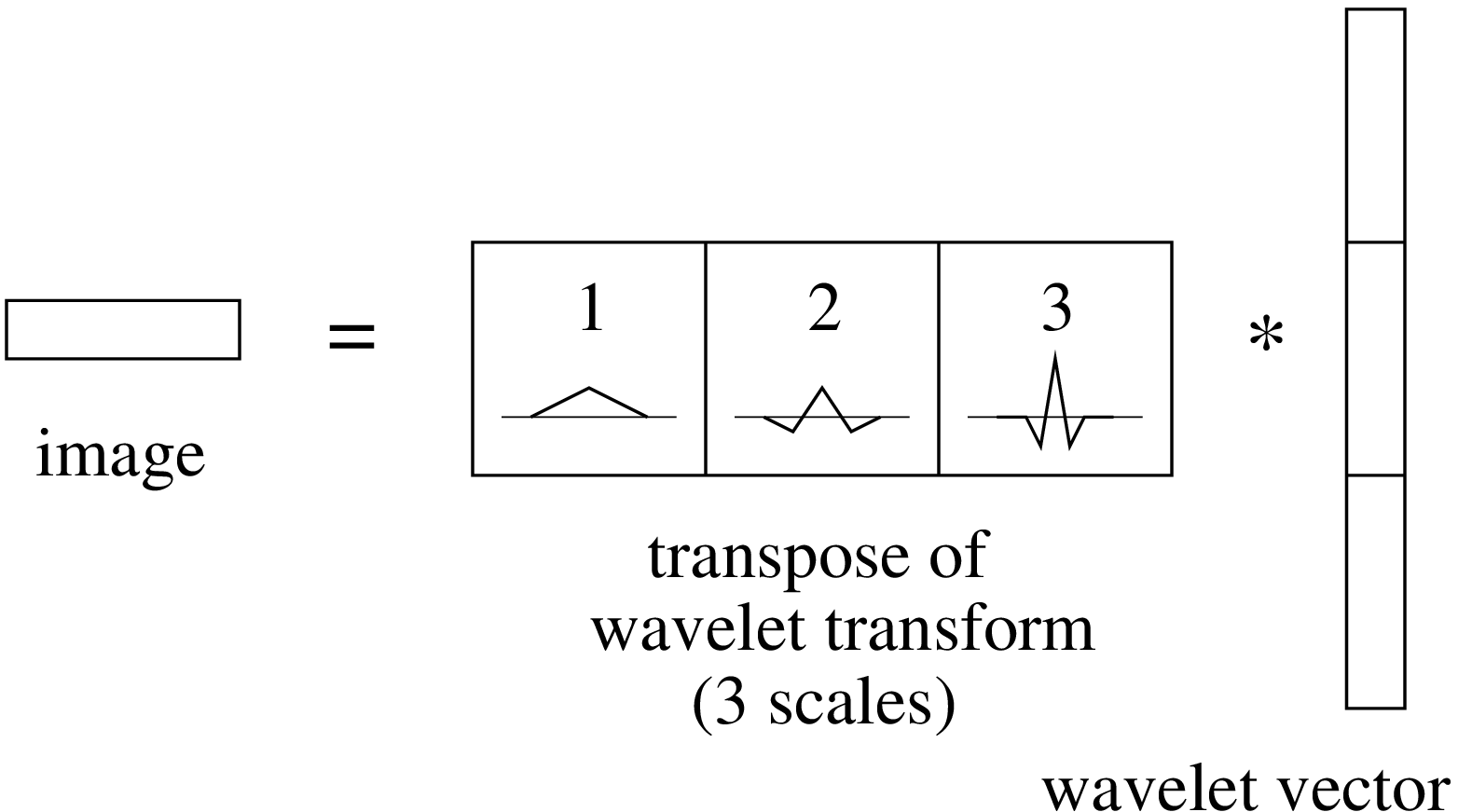}
    \end{tabular}
    \caption{(a) The operation of a 3-level \`{a} trous transform on an
      image.  (b)~The transpose of the \`{a} trous transform operates
      on a wavelet vector with 3 scales.  Each scale is convolved with
      the corresponding wavelet, and the resulting vector is obtained
      by a sum over all scales. }
    \label{fig:atroustranspose}
  \end{center}
\end{figure}

From~(\ref{eqn:atrousconvolution}), the \`{a} trous wavelet
coefficients~$w^j_i$ at the $j$th scale are given by the (discretised)
convolution
\begin{equation}
w^j_i = \sum_k W^j_{ik} f_k = \sum_k \psi^j_{k-i} f_k,
\label{eqn:atrouscoeff}
\end{equation}
of the original function or image~$\myvec{f}$ with a version of the
wavelet~$\psi^j$ at the $j$th scale that was mirrored at the origin.
For a symmetric wavelet function, this corresponds to a convolution
with the wavelet itself.  Examples of wavelet functions are shown in
Figs.~\ref{fig:atrousbasis1} and~\ref{fig:atrousbasis2}.  (For an
orthogonal wavelet the equivalent to~(\ref{eqn:atrouscoeff}) would be
$
w_i = \sum_k \psi_{k-2i} h_k,
$
where the factor of~2 in the index accounts for the decimation carried
out between consecutive filter steps.)  The transpose of the \`{a}
trous transform operates on the wavelet coefficients and produces a
new image~$\myvec{g}$, which is given by
\begin{equation}
g_k = \sum_{j,i} W^j_{ki} w^j_i = \sum_{j,i} \psi^j_{k-i} w^j_i.
\label{eqn:atroustransp}
\end{equation} 
Thus the transpose consists of a convolution of the $j$th wavelet
domain with the corresponding wavelet~$\psi^j$ and a subsequent
summation over all scales.  The operation of the \`{a} trous transform
and of its transpose is illustrated in
Fig.~\ref{fig:atroustranspose}.
\begin{figure*}
  \begin{center}
    \leavevmode
    \begin{center}
      \begin{tabular}{ll}
        \includegraphics[angle=-90,width=7.1cm]{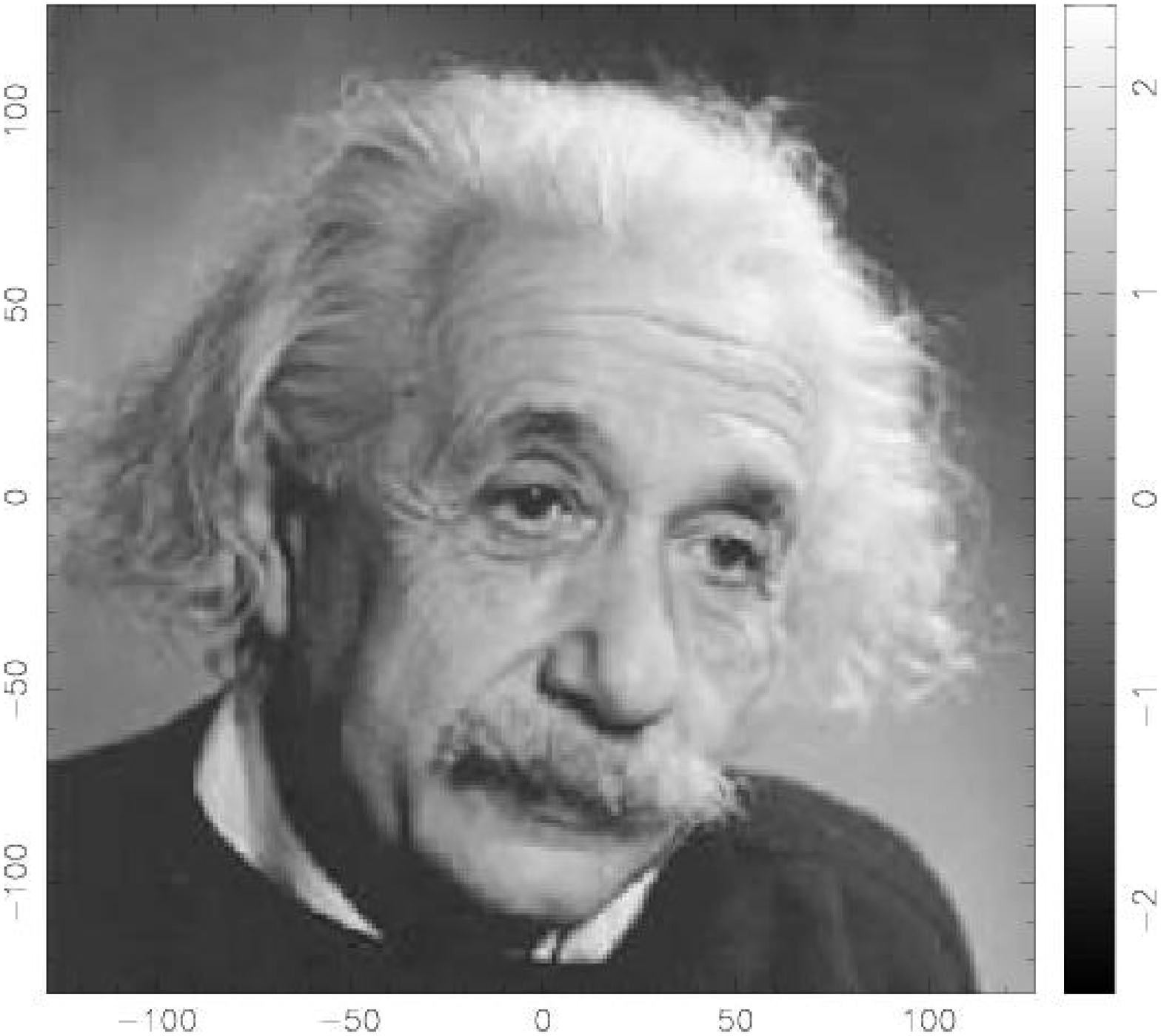}&
        \includegraphics[angle=-90,width=7.1cm]{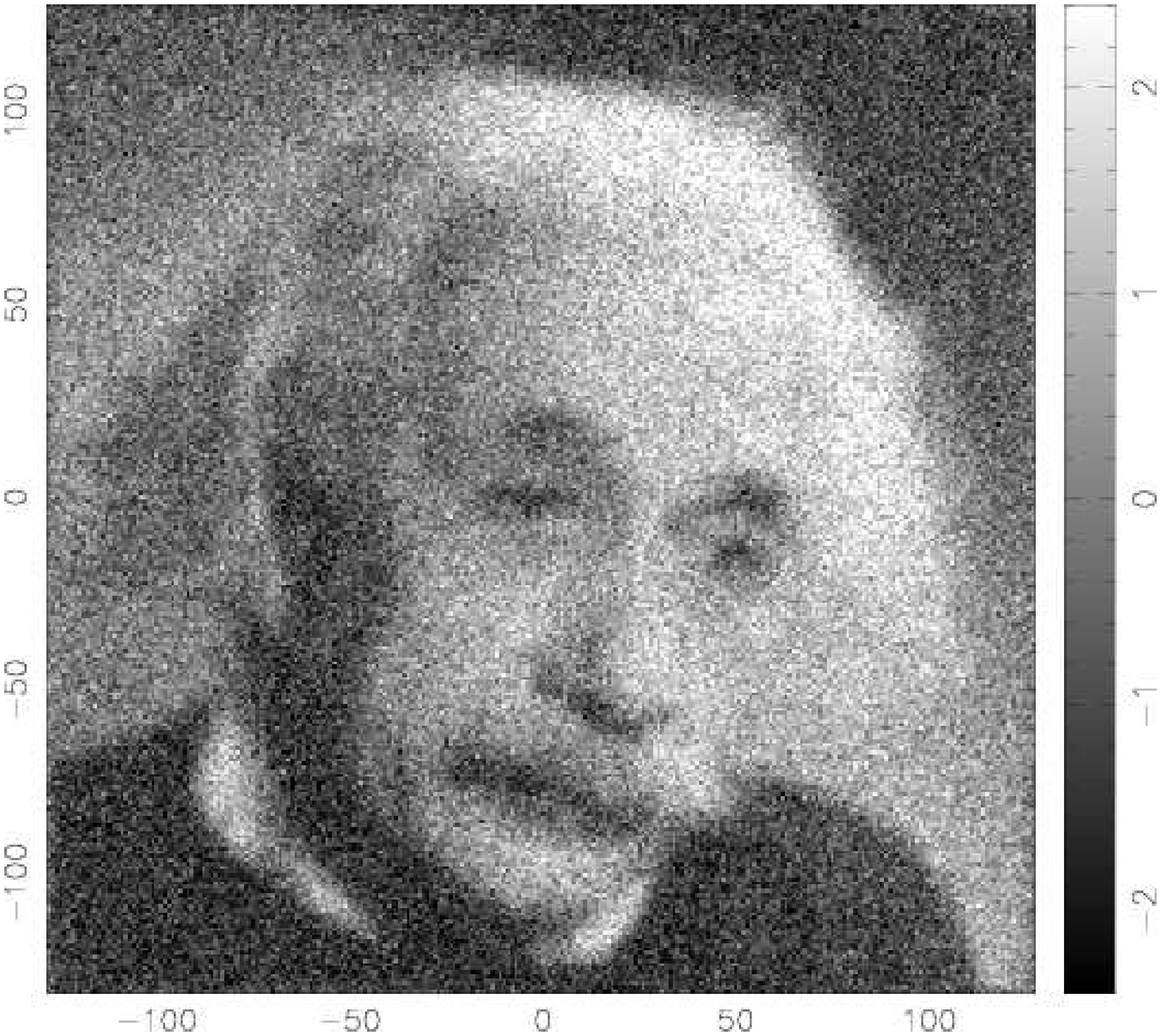}\\
        (a) & (b)
      \end{tabular}
    \end{center}
    \caption{The original image~(a) used in the simulations described in
    the text and the `data image'~(b) obtained after a convolution with
    a Gaussian point spread function with a FWHM of 5~pixels and the
    addition of Gaussian random noise whose \rms\ is half the \rms\ of
    the original image.}
    \label{fig:einstein:data}
  \end{center}
\end{figure*}

The effect of the transpose of the \`{a} trous transform is just the
same as that of a multi-channel ICF.  Its set of ICFs consists of the
scaling function and the rescaled wavelet functions.  Unlike in the
standard multi-channel approach, the blurring functions take negative
values in some areas.  As in Section~\ref{sec:memwavelet:multichannel},
one can introduce different weights for all channels in the \`{a}
trous transform, or one can simply adapt the default model on
different scales to enhance the reconstruction quality.  However, if
one introduces scale-dependent weights~$w_k$, the inverse transform
also needs to be rescaled.  The nice property that the Fourier
transforms of all wavelet functions add up to~1 is lost.

%
%

\section{Application to simulated data}
\label{sec:memwavelet:application}

In this section, we compare the different maximum entropy methods
discussed in this paper by applying them to a set of simulated
two-dimensional images.  One of the problems of assessing the
capabilities of different reconstruction methods is that there is no
single unique criterion for the quality of a reconstruction.
Furthermore, the outcome of any quality measurement depends on a large
number of variables, such as the type and content of the image,
properties of the instrument with which the data were observed (e.g. 
the point spread function and the noise) and the model and
regularisation constant used in the maximum entropy algorithms etc.
In this section, we use the \rms\ differences between the original
image and the reconstruction as a measure of the reconstruction
quality.  Of course, in real applications one does not have the
original image available for comparison, but nevertheless one would
usually hope to minimise the expected difference between true and
reconstructed image.  In
Section~\ref{sec:memwavelet:application:einstein} we use a photographic
image as an arbitrarily chosen, but fairly general and typical test
case for image reconstructions.  In
Section~\ref{sec:memwavelet:application:cmb} we then turn to the
investigation of the reconstruction of CMB~maps that are realisations
obtained from an inflationary CDM model, and provide a useful
astronomical test image that is complementary in its properties to the
photographic image.

\subsection{A photographic image}
\label{sec:memwavelet:application:einstein}

As a first test image we use the photographic portrait shown in
Fig.~\ref{fig:einstein:data}~(a).  The image contains $256 \times 256$
pixels and has had its mean subtracted and has been rescaled to an \rms\ 
of~1.  It is highly non-gaussian, and its sharp lines and contrasting
continuous extended regions provide useful characteristics to assess
the visual impression of different reconstruction methods.  We
simulate observations of this image by convolving it with different
Gaussian point spread functions with FWHMs of 3, 5 and 10~pixels and
adding Gaussian white noise with an \rms\ of 0.1, 0.5 or~2.  Bearing
in mind that the original image had unity \rms, this corresponds to
signal-to-noise ratios of 10, 2 and 0.5.  For each FWHM and noise
level, we use Monte-Carlo simulations of 15~different noise
realisations.  One of the realisations obtained for a FWHM of 5~pixels
and a noise level of~0.5 is shown in Fig.~\ref{fig:einstein:data}~(b).

\begin{figure}
  \begin{center}
    \leavevmode
    \includegraphics[angle=-90,width=8cm]{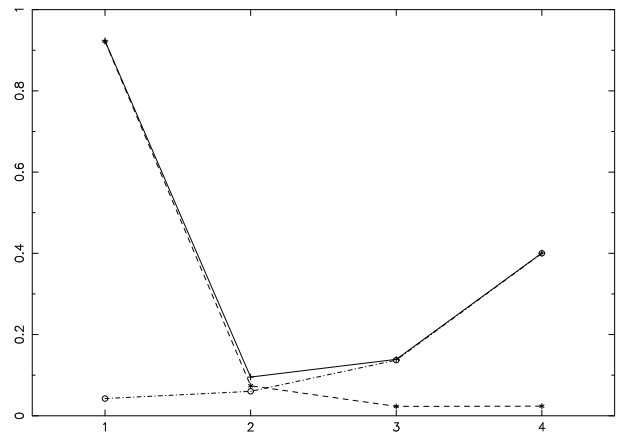}
    \caption{The dispersion of the wavelet coefficients of data (solid
      line), the signal (long-dashed line) and noise contribution
      (short-dashed line) of the image in
      Fig.~\ref{fig:einstein:data}~(b).  There are four levels in the
      \`{a} trous transform, where the level~1 corresponds to the
      coarse structure and level~4 is the domain with the most
      detailed structure. }
    \label{fig:dispersion}
  \end{center}
\end{figure}
\begin{figure}
  \begin{center}
    \leavevmode
    \includegraphics[width=8cm,height=8cm,angle=-90]{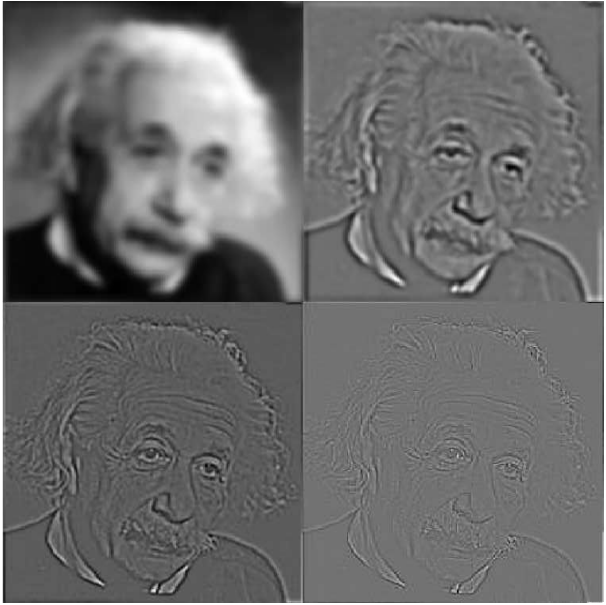}
    \caption{The 4-level \`{a} trous transform of the image in Fig.~\ref{fig:einstein:data}~(a).  Each level is plotted on a different greyscale.}
    \label{fig:atrous:einstein}
  \end{center}
\end{figure}
To each simulation, we apply several different reconstruction
algorithms.  First, we use a real-space MEM algorithm whose image model
is set to be constant across the image plane to the value of the
estimated signal \rms.  The regularisation
parameter~$\alpha$ discussed in Section~\ref{sec:intro:maxent:alpha} is
chosen from the historic MEM criterion which demands that the
$\chi^2$-statistic for the final reconstruction equal the
number~$N_\mathcal{D}$ of data points, in this case $256\times 256$.
We also apply wavelet regularised MEM and wavelet MEM implementations,
each with Daubechies-4 tensor, Daubechies-4 MRA and \`{a} trous
wavelets.  The \`{a} trous wavelets use the triangle
function~(\ref{eqn:triangle}) with four levels.  In all cases, the
model is chosen to be constant within a given wavelet domain.  The
model values for each domain or scale are found by setting them to
the expected signal dispersions as discussed in
Section~\ref{sec:memwavelet:alphamodel}.  Fig.~\ref{fig:dispersion} plots
data, signal and noise dispersions for the 4-level \`{a} trous
transform.  One can clearly see that on coarse scales (level~1) the
data are dominated by the signal and on the finest scales (level~4) by
the noise.  The 4~levels of the \`{a} trous transform are shown in
Fig.~\ref{fig:atrous:einstein}.  Each level is plotted on its own
greyscale.  The detail levels would be much fainter if they were
plotted on the same scale.

\begin{figure*}
  \begin{center}
    \leavevmode
    \begin{center}
      \begin{tabular}{ll}
        \includegraphics[angle=-90,width=7.1cm]{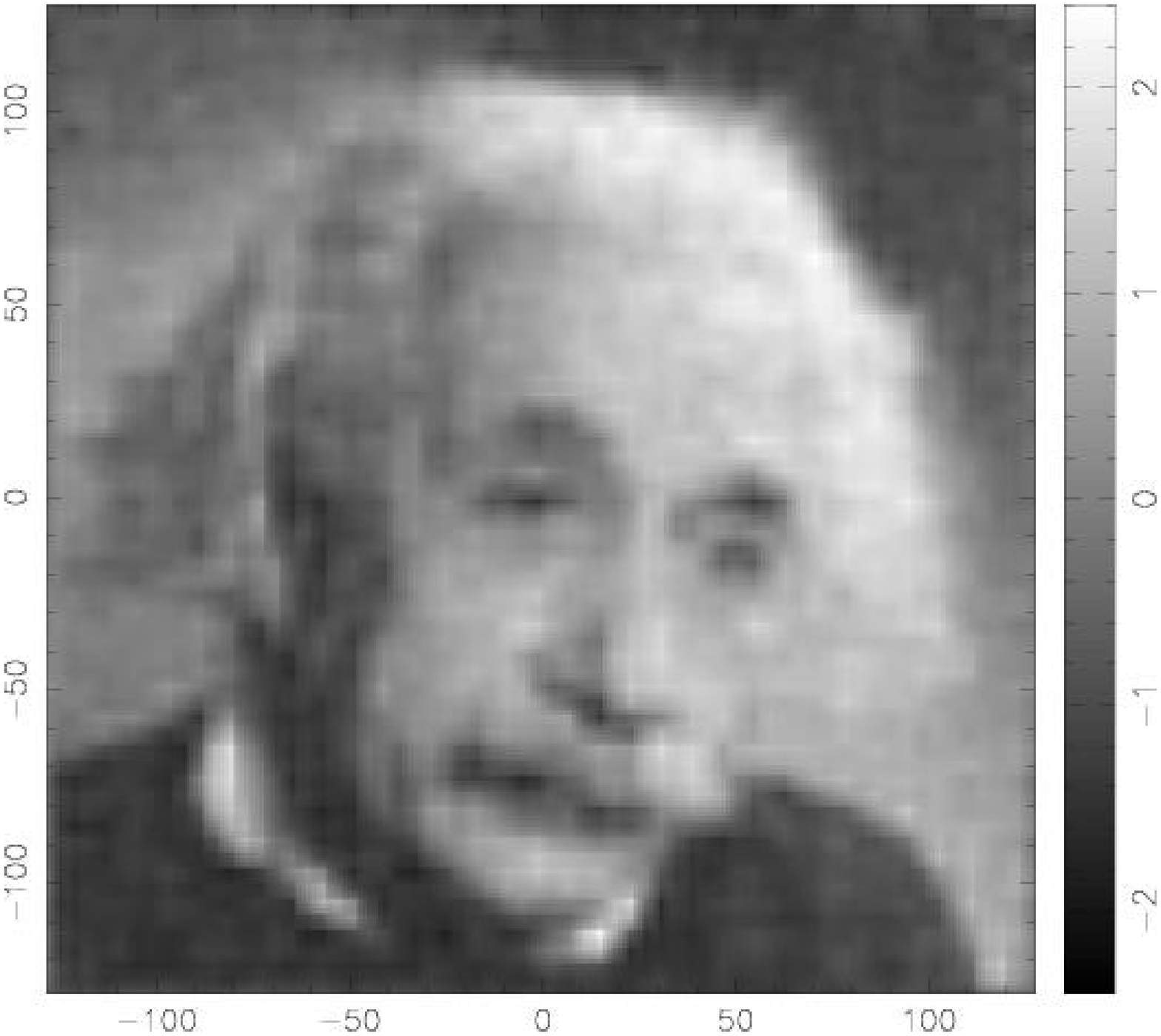}&
        \includegraphics[angle=-90,width=7.1cm]{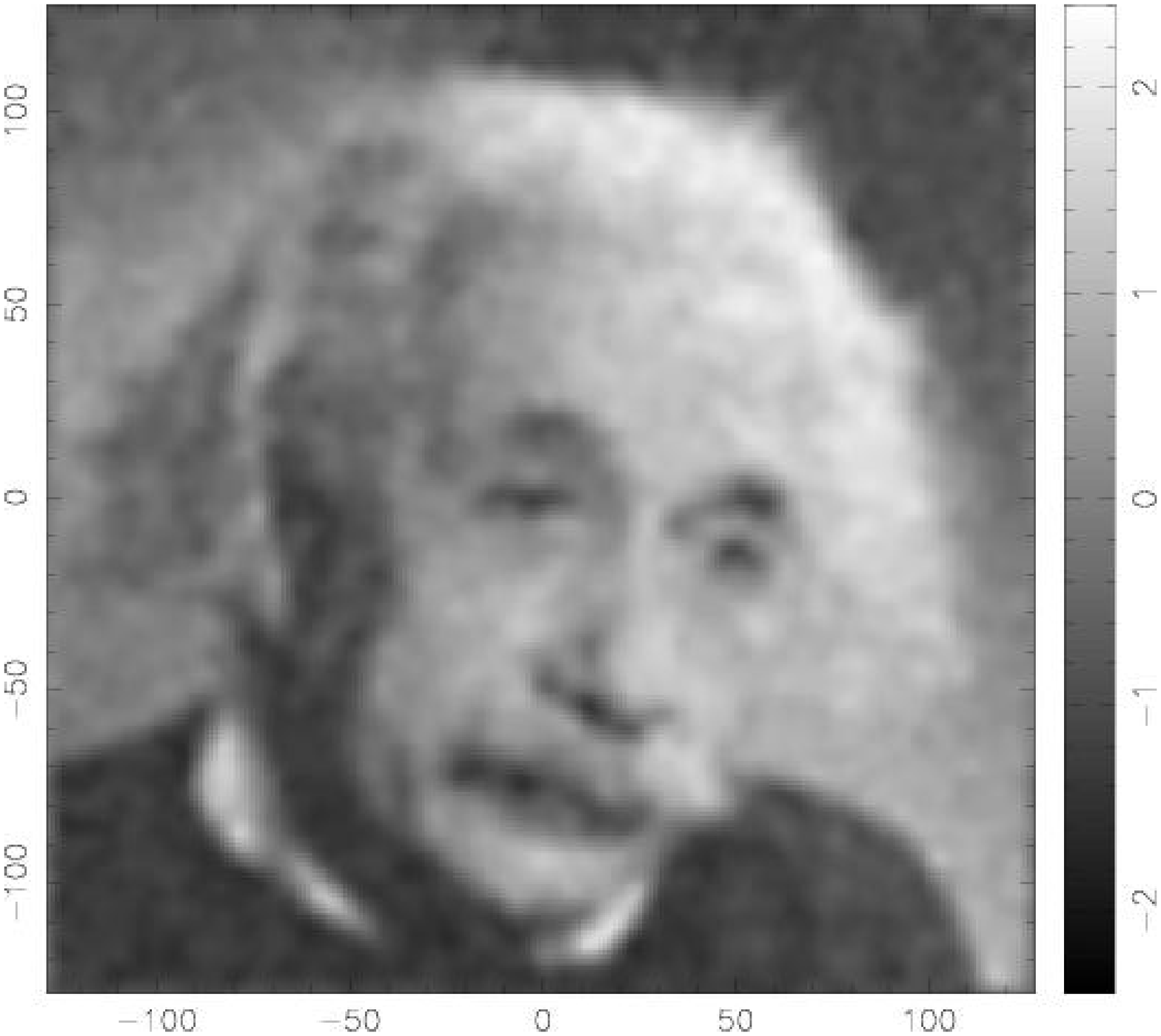}\\
        (a) & (b)\\
        \includegraphics[angle=-90,width=7.1cm]{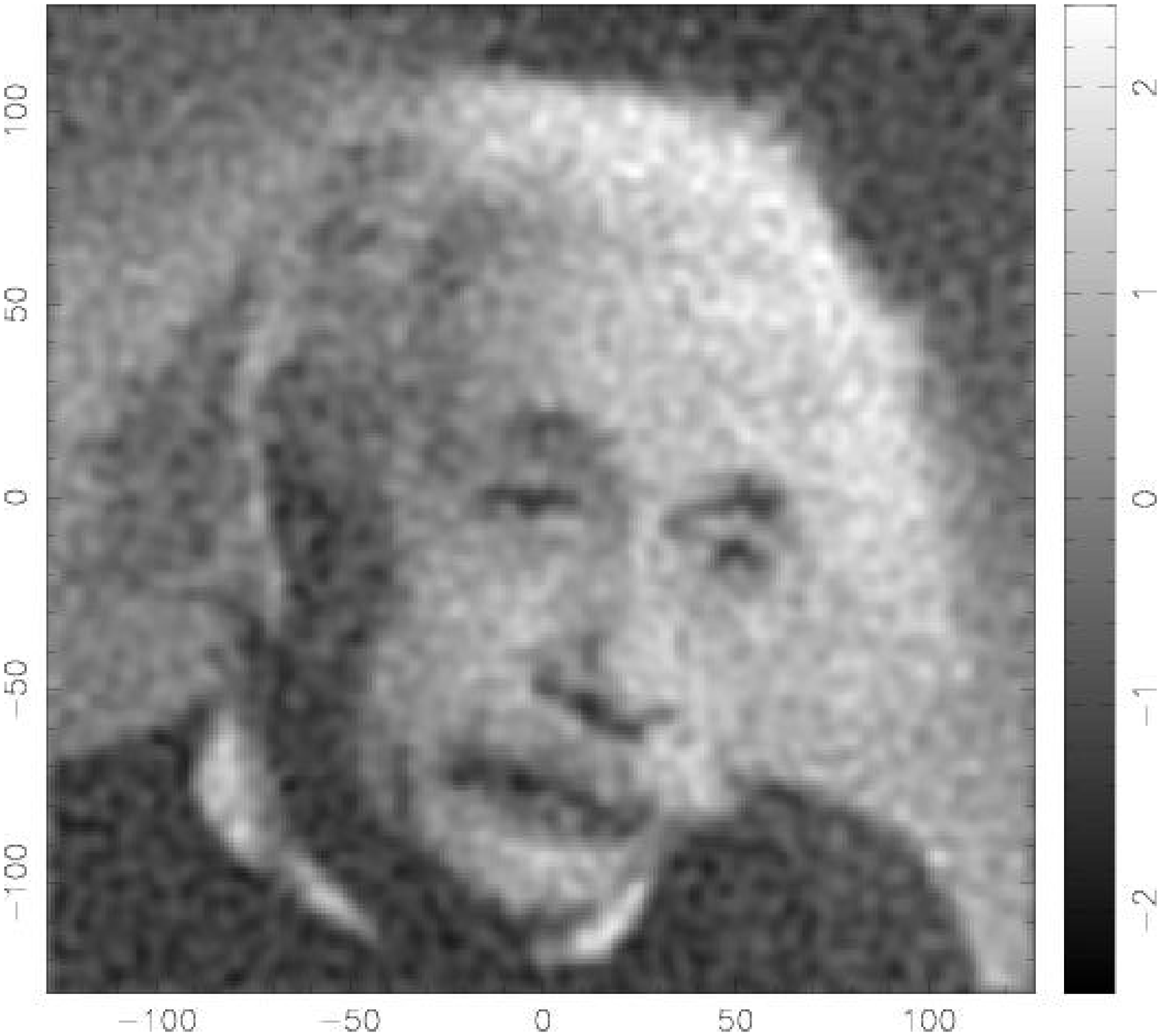}&
        \includegraphics[angle=-90,width=7.1cm]{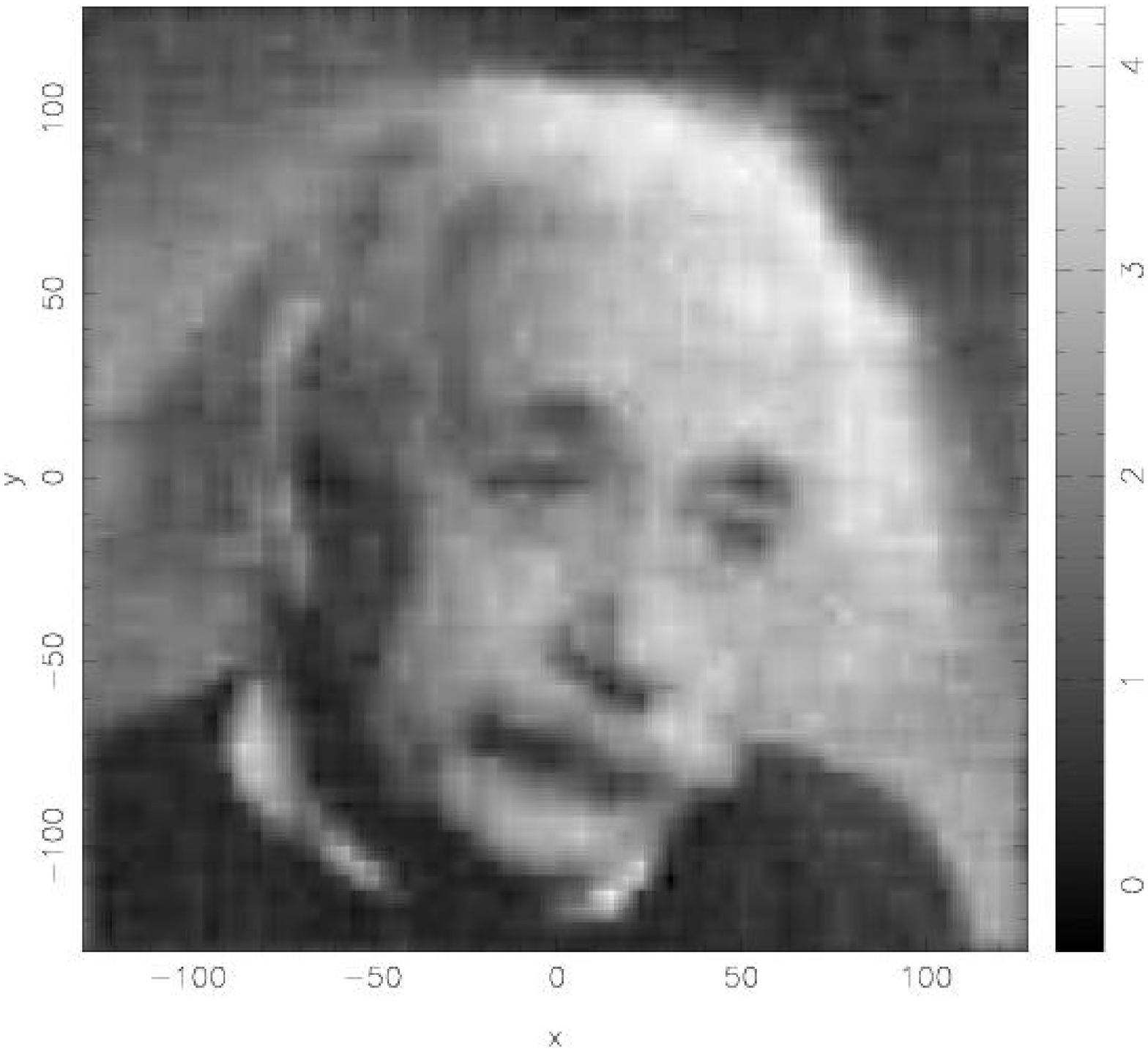}\\
        (c) & (d)
      \end{tabular}
    \end{center}
    \caption{Reconstructions of the data
      from~Fig.~\ref{fig:einstein:data}~(b) using the `wavelet
      regularised MEM' algorithm: (a)~with Daubechies-4 tensor
      wavelets; (b) or with 4-level \`{a} trous wavelets.  A
      reconstruction with real-space MEM is shown in~(c), and a
      \textit{SureShrink} reconstruction in~(d).  Compare the original
      image in Fig.~\ref{fig:einstein:data}~(a).  }
    \label{fig:einstein:reg}
  \end{center}
\end{figure*}

Some of the reconstructions produced from the `data' in
Fig.~\ref{fig:einstein:data}~(b) are presented in
Fig.~\ref{fig:einstein:reg}.  The first image~(a) shows a
reconstruction obtained with the wavelet regularised MEM using tensor
wavelets.  The image looks more or less smooth.  However, there are
some faint structures along the horizontal and vertical directions.
These are the signatures of the highly non-differentiable Daubechies-4
wavelets (compare Fig.~\ref{fig:haardaub}~(b) for the one-dimensional
profile).  The second image~(b), which was produced with the \`{a}
trous wavelets, is the visually most appealing reconstruction.  The
image is very smooth, but due to the high noise levels on the data
there is no apparent attempt at a deconvolution of the point spread
function.  In~(c) the results from real-space MEM are shown.  The
image is dominated by the `ringing' or little speckles that are
characteristic for the real-space method.  Visually, this image is
clearly the poorest reconstruction of the MEM reconstructions.
Finally, in~(d) we show a reconstruction obtained from a
\textit{SureShrink} filter with Daubechies-4 tensor wavelets, as
introduced in Section~\ref{sec:memwavelet:filter}.  As in~(a), the spiky
signatures of the Daubechies-4 wavelets are visible in the
image.

\begin{figure*}
  \begin{center}
    \leavevmode
    \begin{center}
      \begin{tabular}{ll}
        \includegraphics[angle=-90,width=7.1cm]{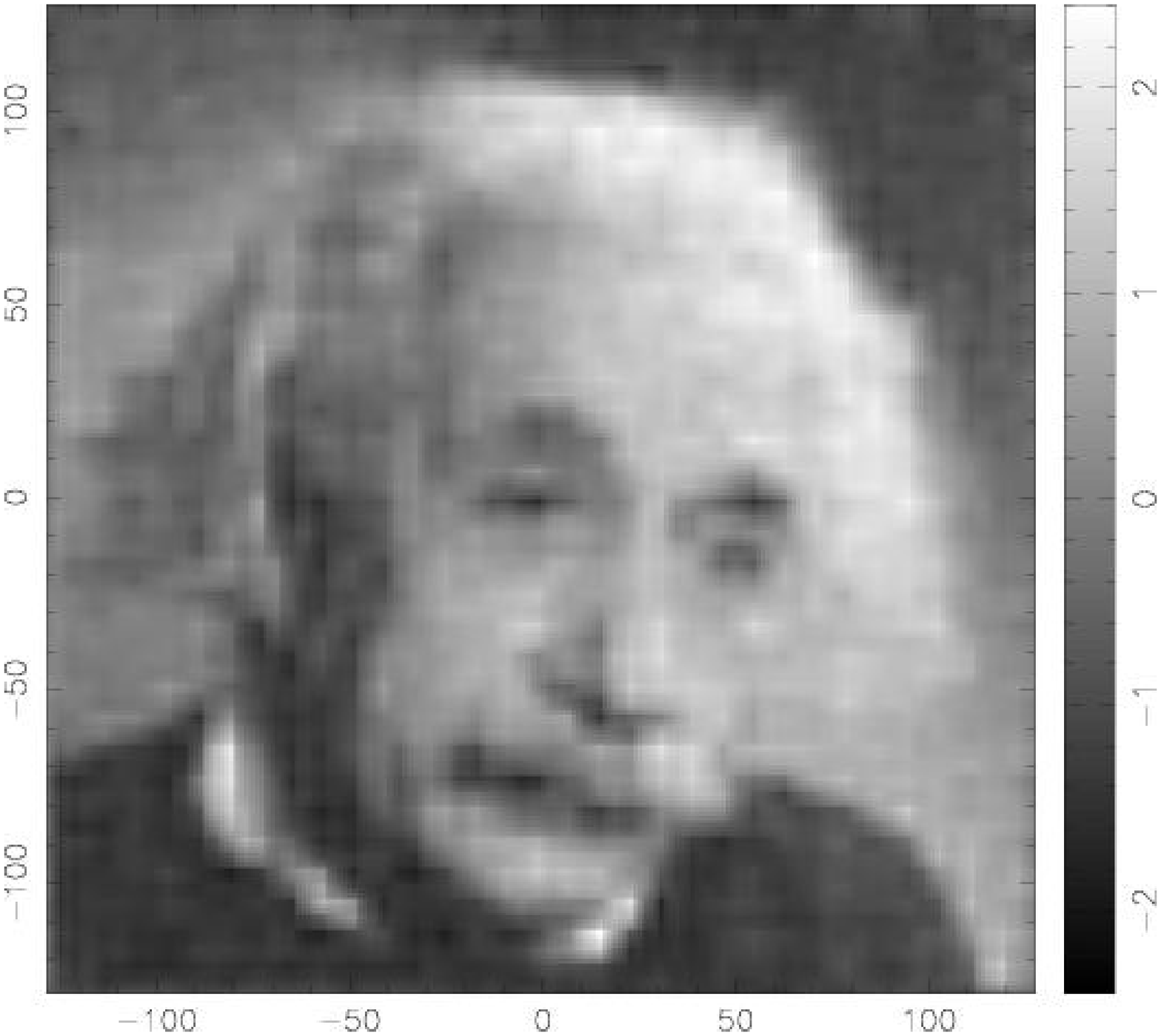}&
        \includegraphics[angle=-90,width=7.1cm]{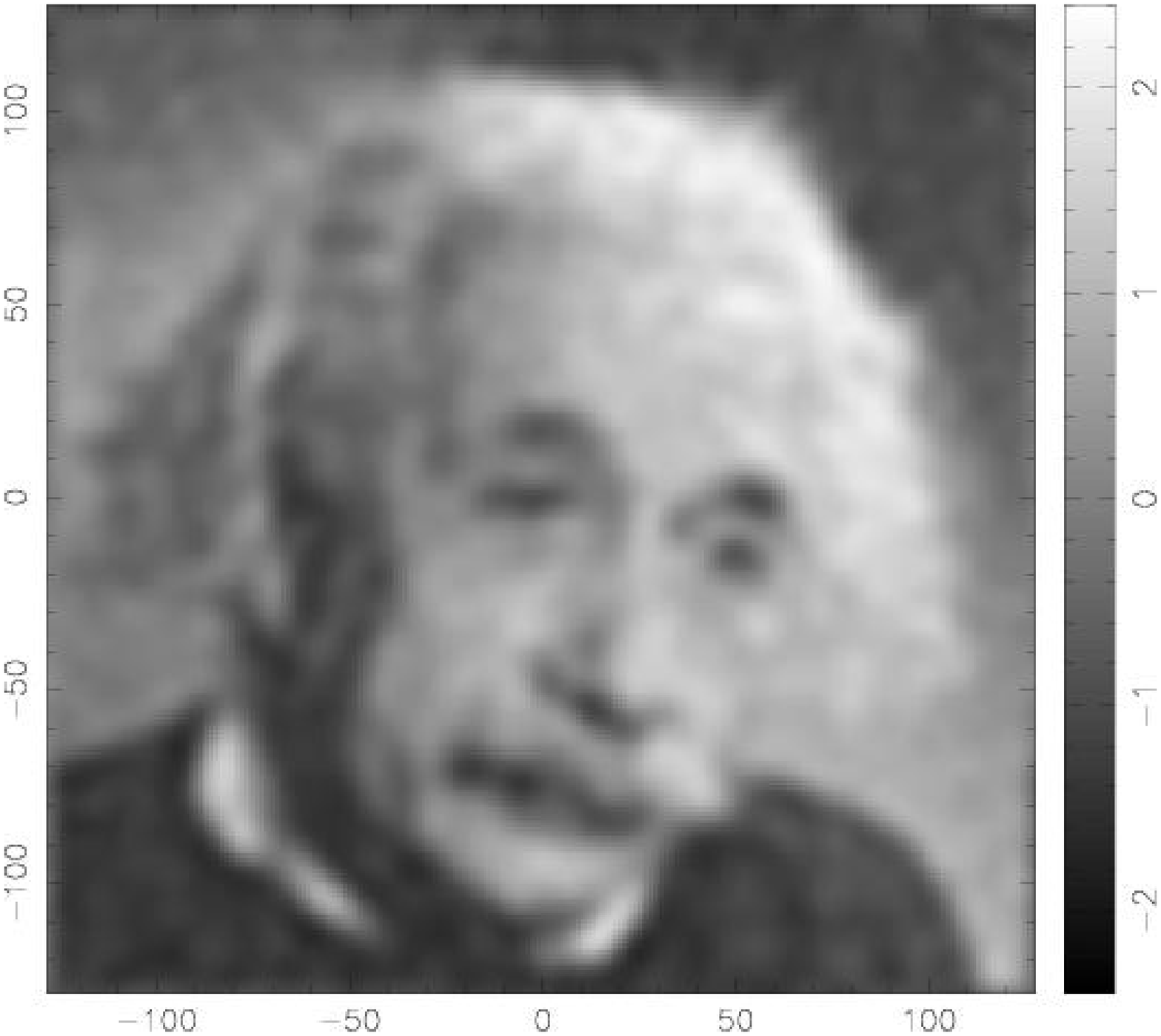}\\
        (a) & (b)
      \end{tabular}
    \end{center}
    \caption{Reconstructions of the simulated image from
      Fig.~\ref{fig:einstein:data} with `wavelet MEM': (a)~ using
      Daubechies-4 tensor wavelets; (b) using \`{a} trous wavelets.
      Compare the corresponding reconstructions obtained from `wavelet
      regularised MEM' shown in Figs.~\ref{fig:einstein:reg}~(a)
      and~(b).}
    \label{fig:einstein:wave}
  \end{center}
\end{figure*}
The reconstructions for the wavelet MEM using tensor and \`{a} trous
wavelets are shown in Fig.~\ref{fig:einstein:wave}.  They look very
similar to the results obtained from the wavelet regularised MEM in
Fig.~\ref{fig:einstein:reg}.

\begin{table*}
  \begin{center}
    \leavevmode
    \begin{tabular}{c||ccc|ccc|ccc}
      \hline \hline
      PSF {FWHM} & \multicolumn{3}{c|}{3} &\multicolumn{3}{c|}{5}
         &\multicolumn{3}{c}{10} \\
      noise \rms & 0.1 & 0.5 & 2.0 & 0.1 & 0.5 & 2.0 & 0.1 & 0.5 & 2.0 \\
         \hline
         real-space MEM& 0.12 & 0.30 & 0.60 & 0.14 & 0.25 & 0.49 & 0.21 &
         0.26 & 0.41\\[0.5ex]
         tensor reg. MEM & 0.12 & 0.20 & 0.39 & 0.15 & 0.22 & 0.37 &
         0.21 & 0.26 & 0.36\\
         \`{a} trous reg. MEM & 0.11 & 0.21 & 0.45 & 0.14 &
         0.21 & 0.39 & 0.21 & 0.26 & 0.37\\[0.5ex]
         tensor wavelet MEM & 0.12 & 0.20 & 0.34 & 0.16 & 0.22 & 0.34 &
         0.22 & 0.27 & 0.36\\
         \`{a} trous wavelet MEM & 0.11 & 0.18 & 0.32 & 0.14 & 0.21 &0.32
         & 0.21 & 0.25 & 0.35\\[0.5ex]
         \textit{SureShrink} & 0.12 & 0.19 & 0.32 & 0.17 & 0.22 & 0.36
         & 0.26 & 0.29 & 0.36 \\ \hline \hline
    \end{tabular}
    \caption{Reconstruction errors for different methods as a
      function of the FWHM (in image pixels) of the convolution mask
      and of the noise level on the data.  The numbers quote the \rms\
      differences between the reconstruction and the original
      image averaged over a number of noise realisations as
      described in the text.}
    \label{tab:memwavelet:rec:einstein}
  \end{center}
\end{table*}
For a more quantitative comparison of the different methods,
Table~\ref{tab:memwavelet:rec:einstein} lists the reconstruction
errors for different FWHMs of the point spread function (3, 5 and 10
pixels) and different noise levels (0.1, 0.5 and 2).  The quoted
errors are the \rms\ differences between the original image and the
reconstruction averaged over 15 different noise realisations.  The
standard deviation of the error values is usually much less than 0.01
for low noise values and not more than 0.03--0.04 in the high-noise
case, so the error on the mean (of the errors) is expected to be not
more than 0.01 in most cases.  There are several trends apparent from
the numbers:
\begin{itemize}
\item For large point spread functions, all methods yield similar
  results (except perhaps for the real-space MEM at low
  signal-to-noise).
\item For high signal-to-noise, the methods also have a similar
  performance.
\item For sufficiently narrow point spread functions and poor
  signal-to-noise ratios, real-space MEM
  performs clearly worse than its competitors.
\item There is some indication that the wavelet MEM performs better
  than the wavelet regularised MEM. 
\item Also, there are some hints that \`{a} trous wavelets may
  outperform the tensor wavelets at least for the wavelet MEM,
  although in wavelet regularised MEM they are less able to cope with
  high noise levels.
\item In the low signal-to-noise regime the \textit{SureShrink} filter
  matches the performance of the wavelet MEM techniques, indicating
  that the errors are entirely dominated by the noise.  For low noise
  levels, however, MEM is more effective than the filter, which makes
  no attempt at a deconvolution.
\end{itemize}
In the case when the point spread function is narrow and the noise
dominant, it is most difficult for the MEM algorithms to distinguish
between signal and noise and the performance differences become most
prominent.

Within the wavelet algorithms, there are of course many free
parameters that may influence the reconstruction errors.  We have
performed simulations with different types of orthogonal wavelets and
both two-dimensional tensor and MRA transforms.  There is no
significant difference except for the case of the simple Haar
wavelets, which seem to perform slightly worse than other wavelets.
Likewise, for the \`{a} trous algorithm the triangle function seems to
be similarly efficient as the $B_3$-spline, and the number of levels
in the transform does have marginal effects as long it is greater than
a minimum of three or four.  We have also tested different models for
the wavelet coefficients.  Generally, there are many models
suppressing power on small wavelet scales that provide a similar
performance.  A stronger relative penalty on small-scale structure,
for example by using the variance instead of the dispersion of the
signal, can be advantageous for poor signal-to-noise ratios, since
most of the fine structure will be noise.  We find no benefits from
methods that attempt a more data-dependent regularisation, for
instance using the regularisation constant~(\ref{eqn:psalpha}).

Tracing back the reconstruction errors to the wavelet domain, it
appears that, not surprisingly, the dominant contribution to the
errors comes from those domains where the signal-to-noise is close to
unity.  The other domains are either perfectly reconstructed (the
coarse structure) or contribute little to the image (the
noise-dominated small-scale structure).

%
%

\subsubsection{The influence of the regularisation constant~$\alpha$}
\label{sec:memwavelet:application:alpha}

\begin{figure}
  \begin{center}
    \leavevmode
    \begin{center}
      \includegraphics[angle=-90,width=8cm]{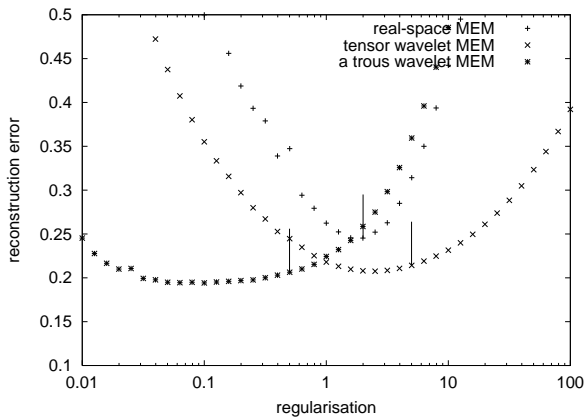}
    \end{center}
    \caption{The \rms\ reconstruction errors as a function of the
      regularisation constant~$\alpha$, for a simulation with a
      5-pixel FWHM blurring function and a signal-to-noise of~2.  For
      real-space MEM (asterisks) and wavelet MEM using tensor
      (diagonal crosses) and \`{a} trous wavelets (perpendicular
      crosses), the plotted points show the final errors for a
      reconstruction with a fixed value of~$\alpha$.  For each method,
      the small lines indicate the value of~$\alpha$ that would have
      been obtained from the historic MEM criterion $\chi^2 =
      N_\mathcal{D}$. }
    \label{fig:memwavelet:alpha}
  \end{center}
\end{figure}
Beside the choice of the model, the selection of the regularisation
constant~$\alpha$ is one of the major problems in the maximum entropy
method, as already discussed in Sections~\ref{sec:intro:maxent:alpha}
and \ref{sec:memwavelet:alphamodel}.  By setting the
$\chi^2$-statistic to the number of data points, $\chi^2 =
N_\mathcal{D}$, one essentially fixes the `goodness of fit', which may
result in very similar error levels on the reconstructions independent
of the basis functions or regularisation.
Fig.~\ref{fig:memwavelet:alpha} illustrates the dependence of the
reconstruction error on the regularisation constant~$\alpha$ for
real-space MEM and wavelet MEM using tensor and \`{a} trous wavelet.
These results are obtained from a simulation with a FWHM of 5~pixels
and a noise level of~0.5.  Each point corresponds to the \rms\ 
difference between the final reconstructions and the original image
when the regularisation parameter~$\alpha$ has been fixed to the given
value.  One can clearly see that there is a trade-off curve between
too close a fit to the data (and thus to spurious noise) and too
strong a regularisation (and thus to suppression of real structure in
the image).  Both extremes result in high reconstruction errors.  For
each reconstruction method, a short horizontal line marks the point
along the curve that is preferred by the $\chi^2=N_\mathcal{D}$
criterion.  Two features of these curves are worth pointing out:
\begin{itemize}
\item The global minima of the curves are indeed lower for the wavelet
  methods than for real-space MEM.  This means that the lower errors
  of the wavelet methods in Table~\ref{tab:memwavelet:rec:einstein}
  are not merely artefacts of a poorly chosen regularisation constant.
  On the contrary, for the real-space algorithm the historic MEM
  criterion actually picks out points that are very close to the
  global minimum of the curve, despite the visual ringing evident from
  the image.
\item The curves appear to be less narrowly peaked for the wavelet
  methods than for real-space MEM.  This means the quality of the
  reconstruction will be less sensitive to~$\alpha$.  For the wavelet
  methods, one can vary $\alpha$ by about an order of magnitude around
  the minimum without significantly affecting the reconstruction
  errors.
\end{itemize}

\subsubsection{A Bayesian choice of the regularisation constant~$\alpha$}
\label{sec:memwavelet:application:bayes}

\begin{table*}
  \begin{center}
    \leavevmode
    \begin{tabular}{c||ccc|ccc|ccc}
      \hline \hline
      PSF {FWHM} & \multicolumn{3}{c|}{3} &\multicolumn{3}{c|}{5}
         &\multicolumn{3}{c}{10} \\
      noise \rms & 0.1 & 0.5 & 2.0 & 0.1 & 0.5 & 2.0 & 0.1 & 0.5 & 2.0 \\
         \hline
         real-space MEM& 0.47 & 0.63 & 0.85 & 0.43 & 0.57 & 0.74 &
         0.38 & 0.50 & 0.65\\[0.5ex]
         tensor wavelet MEM & 0.12 & 0.21 & 0.35 & 0.15 & 0.23 & 0.35
         & 0.22 & 0.27 & 0.37\\
         \`{a} trous wavelet MEM & 0.15 &0.27 &0.43 &0.15 &0.22 &0.35 &0.20
         &0.27 &0.42\\[0.5ex]
         ICF MEM & 0.12 & 0.24 & 0.44 & 0.18 & 0.28 & 0.45 & 0.25 &
         0.34 & 0.47\\
         \hline \hline
    \end{tabular}
    \caption{Reconstruction errors as in
      Table~\ref{tab:memwavelet:rec:einstein}, but for a Bayesian
      choice of the regularisation constant~$\alpha$ (classic MEM).}
    \label{tab:memwavelet:rec:einsteinalpha}
  \end{center}
\end{table*}
\begin{table*}
  \begin{center}
    \leavevmode
    \begin{tabular}{c||ccc|ccc|ccc}
         \hline \hline
      PSF {FWHM} & \multicolumn{3}{c|}{3} &\multicolumn{3}{c|}{5}
         &\multicolumn{3}{c}{10} \\
      noise \rms & 0.1 & 0.5 & 2.0 & 0.1 & 0.5 & 2.0 & 0.1 & 0.5 & 2.0\\
      \hline
         real-space MEM&  0  & 0 & 0 & 0 & 0 &
         0 & 0 & 0 &  0\\[0.5ex]
         tensor waveletMEM& 17040 & 6910 & 1830 & 6240 & 3060 &
         1010 & 1350 & 930 & 370\\
         \`{a} trous wavelet MEM& 15310 & 5850 & 1360 & 6090 & 2590
         & 730 & 1290 & 120 & 170\\[0.5ex] 
         ICF MEM& 13900  & 5080 & 1220 & 4450 & 1890 & 560
         & 850 & 430 & 150\\ 
         \hline \hline
    \end{tabular}
    \caption{The logarithm~$\ln \Pr(\myvec{d})$ of the
      evidence~$\Pr(\myvec{d})$ for a classic MEM reconstruction with a
      Bayesian choice of the regularisation constant~$\alpha$ (classic
      MEM). The logarithms are normalised such that they equal zero
      for the real-space MEM for each dataset. }
    \label{tab:memwavelet:evidence}
  \end{center}
\end{table*}

The proper Bayesian way to determine the regularisation
constant~$\alpha$ is to treat it as an additional model parameter and
marginalise the posterior probability over~$\alpha$, i.e.  integrate
over all possible values of~$\alpha$ (see
Section~\ref{sec:intro:maxent:alpha}).  Because the evidence is often
strongly peaked, it is usually sufficient to maximise the evidence to
find the optimal value of~$\alpha$.  The reconstruction errors for MEM
with a Bayesian~$\alpha$ (using \textsc{memsys5}) are shown in
Table~\ref{tab:memwavelet:rec:einsteinalpha}.  It is evident that for
the real-space MEM the reconstruction errors are very poor compared to
those obtained with the historic MEM criterion in
Table~\ref{tab:memwavelet:rec:einstein}.  This is expected from the
curve in Fig.~\ref{fig:memwavelet:alpha}: the historic MEM criterion
picks a value of~$\alpha$ that is close to the minimum of the
trade-off curve.  The reconstruction errors can only get worse for a
different choice of~$\alpha$.  For comparison with
Fig.~\ref{fig:memwavelet:alpha}, the Bayesian value for the real-space
MEM is $\alpha=0.12$.  In fact, it is this poor performance of the
real-space method that historically lead to the introduction of ICFs
and the search for a better image model \citep[e.g.][]{mackay92a}.
The wavelet MEM improves the reconstruction errors dramatically,
especially in the case of the tensor wavelets (for comparison with
Fig.~\ref{fig:memwavelet:alpha}, $\alpha = 0.5$).  The errors even
approach those obtained for the \`{a} trous wavelet MEM in
Table~\ref{tab:memwavelet:rec:einstein}.  The \`{a} trous transform
($\alpha = 0.04$ in Fig.\ref{fig:memwavelet:alpha}) has problems for
narrow point spread functions; it tends to allow too much image
structure on small scales.  However, by reweighting the individual
scales of the transform suitably, the reconstructions can be much
improved.  The optimal weighting factors will be investigated in
future work.

The last row of Table~\ref{tab:memwavelet:rec:einsteinalpha} shows the
reconstruction errors obtained from a Gaussian multi-channel ICF with
four channels.  The widths of the Gaussians are scaled by a factor
of~2 between channels, and the weights are chosen such that the volume
of the ICFs are identical between levels.  Even though the weights are
not specifically adapted to the data, the reconstruction errors are
much lower than for real-space MEM. Again, the reconstructions can be
much improved by the choice of different convolution masks and
weightings.

The evidence for classic real-space and wavelet MEM are presented in
Table~\ref{tab:memwavelet:evidence}.  The values are the logarithms
$\ln \Pr(\myvec{d})$ of the evidence~$\Pr(\myvec{d})$ as introduced in
Section~\ref{sec:bayes}. The values have been normalised such that for
a given FWHM and noise \rms\ the evidence for the real-space MEM
equals~1.  The values presented here thus allow a comparison of the
evidence for a model with ICF and the simple real-space MEM without
ICF.  The standard deviation across the different noise realisations
is roughly 170 (in units of $\ln \Pr(\myvec{d})$).  It is apparent
that the evidence for the ICF MEM and wavelet MEM is significantly
higher than for the real-space MEM, as would be expected from the
improved reconstruction errors.  Furthermore, the lower reconstruction
residuals for the tensor transform are matched by a higher evidence.
From a Bayesian point of view, the ICF model or the wavelet basis are
therefore the favoured models for image reconstructions.
Unfortunately, a similar calculation of the evidence for the wavelet
regularised MEM is not possible, because the software package
\textsc{memsys5} that is used for the reconstruction is limited to
regularisation functionals whose curvature matrix is diagonal.

\subsection{CMB maps}
\label{sec:memwavelet:application:cmb}

Having applied the wavelet MEM techniques to a general image, we now
turn to the reconstruction of CMB maps.  We apply the different methods
to five different CMB maps that are realisations obtained from the
inflationary cold dark matter (CDM) model.  The map size is $16\degr
\times 16\degr$ with $256 \times 256$ pixels, corresponding to a pixel
size of $3.75 \arcmin$.  The maps are convolved with Gaussian beams of
different beam sizes with FWHMs of $11.25\arcmin$ (3 pixels),
$18.75\arcmin$ (5 pixels) and $37.5\arcmin$ (10 pixels) respectively.
Gaussian white noise is added with different signal-to-noise ratios of
$\sigma_\mathrm{S}/\sigma_\mathrm{N} = 10$, 2 and $0.5$.  For each
signal-to-noise ratio, we create five different noise realisations.
With the five different input CMB maps, there are thus 25~different
combinations of sky and noise realisations available for a given FWHM
and noise level.  The maps are reconstructed on the same grid as the
simulated data.  Again we use the historic MEM criterion to determine
the regularisation constant instead of classic MEM, since it produces
sufficiently good reconstruction errors and is also easily applicable to
wavelet-regularised MEM.

\begin{table*}
  \begin{center}
    \leavevmode
    \begin{tabular}{c||ccc|ccc|ccc}
      \hline \hline
      PSF {FWHM} & \multicolumn{3}{c|}{3} &\multicolumn{3}{c|}{5}
         &\multicolumn{3}{c}{10} \\
      noise \rms & 0.1 & 0.5 & 2.0 & 0.1 & 0.5 & 2.0 & 0.1 & 0.5 & 2.0 \\
         \hline 
         real-space MEM& 0.14 & 0.34 & 0.66 & 0.26 & 0.39 & 0.62 & 0.44 &
         0.51 & 0.66\\[0.5ex] 
         tensor reg. MEM & 0.18 & 0.33 & 0.57 & 0.28 & 0.40 &
         0.58 & 0.45 & 0.52 & 0.66\\ 
         \`{a} trous reg. MEM & 0.15 & 0.32 & 0.59 & 0.27 &
         0.39 & 0.59 & 0.45 & 0.52 & 0.65\\[0.5ex] 
         tensor wavelet MEM&  0.18 & 0.33 & 0.56 & 0.30 & 0.41 & 0.59 &
         0.47 & 0.54 & 0.67\\ 
         \`{a} trous wavelet MEM& 0.15 & 0.32 & 0.55 & 0.27 & 0.40 & 0.57
         & 0.45 & 0.52 & 0.65\\[0.5ex]
         \textit{SureShrink} filter& 0.24 & 0.34 & 0.55 & 0.38 & 0.44
         & 0.58 & 0.57 & 0.59 & 0.67\\
         \hline \hline
    \end{tabular}
    \caption{Reconstruction errors between reconstructions and original
      CMB maps.  The errors are averaged over a set of 25~simulations
      with 5~different noise and image realisations.}
    \label{tab:memwavelet:rec:cmb}
  \end{center}
\end{table*}
The reconstruction errors for different methods are quoted in
Table~\ref{tab:memwavelet:rec:cmb}.  Because of the wealth of
information present on all scales in the CMB maps, the errors are
generally larger than for the photographic image (compare
Table~\ref{tab:memwavelet:rec:einstein}).  Although the differences in
reconstruction errors are less conspicuous, the results confirm the
conclusions drawn from the photographic reconstructions.  For low
signal-to-noise ratios and narrow point spread functions, the
wavelet-based methods are superior to real-space MEM.  However, there
is no significant difference between wavelet regularised and wavelet
methods, even though there is still some indication that \`{a} trous
wavelets perform better than orthogonal ones for high signal-to-noise
ratios.  In the last row, we also show that the \textit{SureShrink}
filter again cannot match the performance of maximum entropy for high
signal-to-noise ratios.

\begin{figure*}
  \begin{center}
    \leavevmode
    \begin{center}
      \begin{tabular}{ll}
        \includegraphics[angle=-90,width=7.1cm]{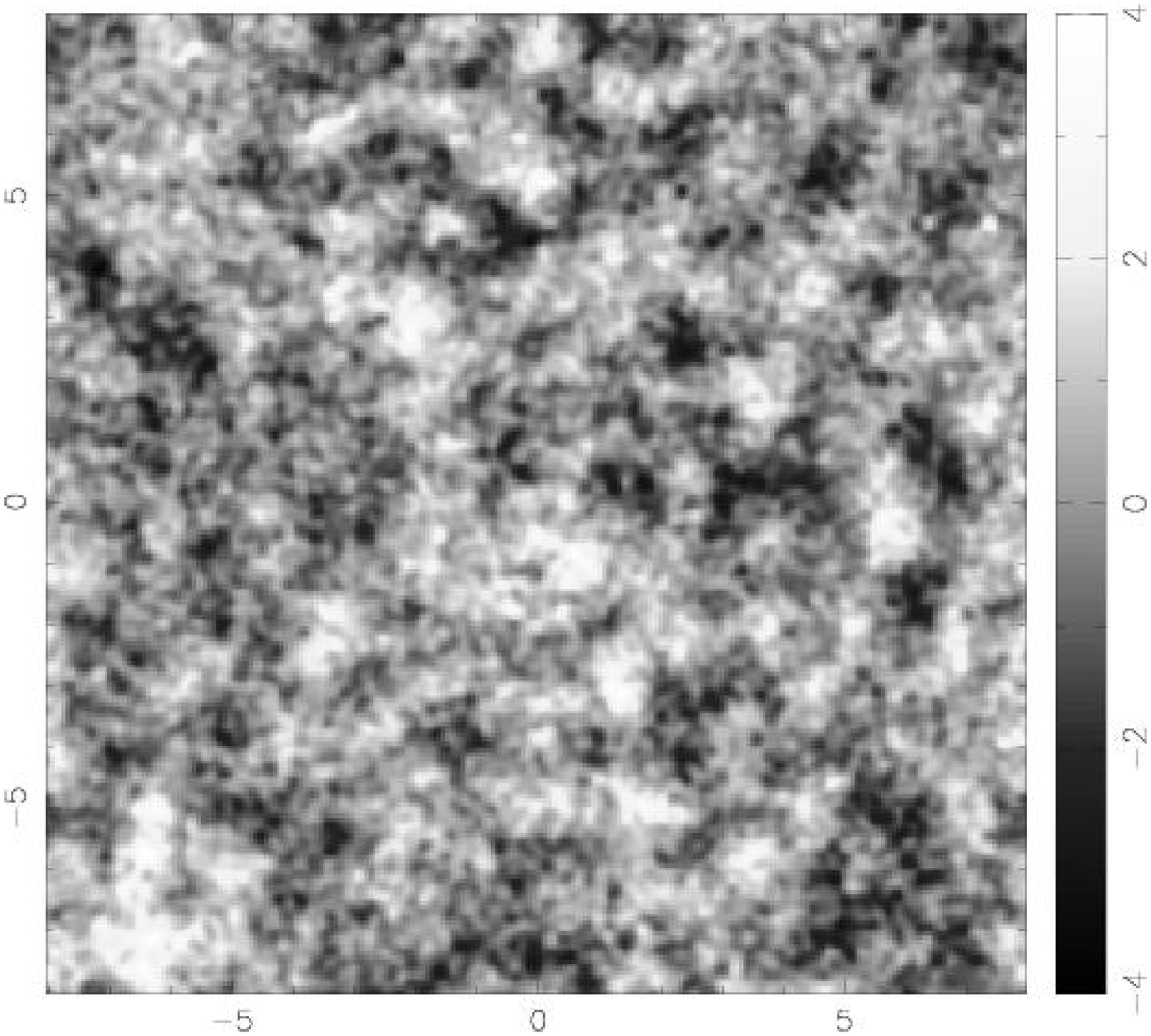}&
        \includegraphics[angle=-90,width=7.1cm]{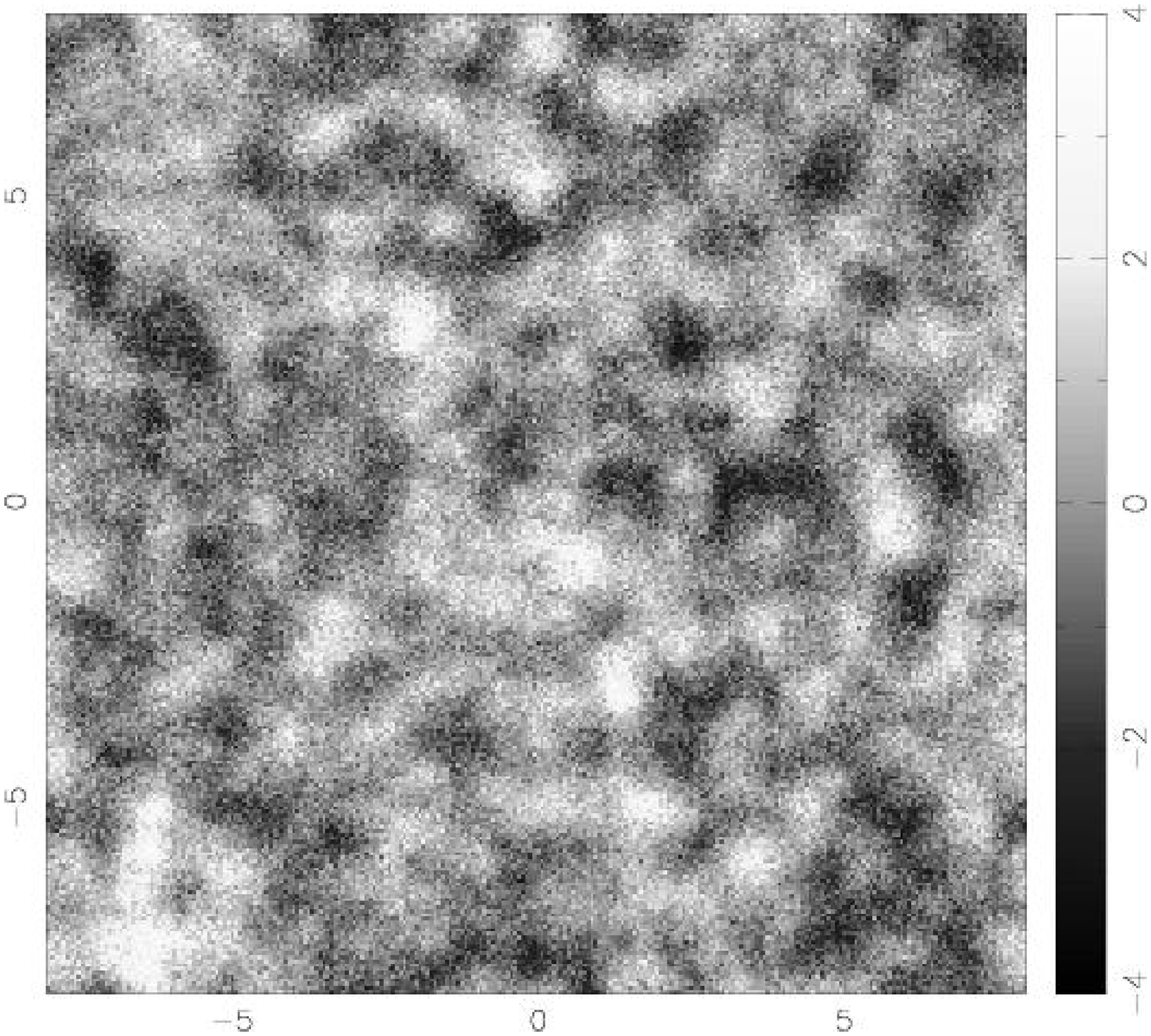}\\
        (a) & (b)
      \end{tabular}
    \end{center}
    \caption{(a)~One realisation of the CMB maps used in the
      simulations described in the text. The map is normalised to an
      unity \rms, and the axes are labelled in degrees. (b)~The `data
      image' obtained after a convolution with a Gaussian point spread
      function with a FWHM of 5~pixels and the addition of Gaussian
      random noise whose \rms\ is half the \rms\ of the original
      image.}
    \label{fig:cmb:data}
  \end{center}
\end{figure*}
\begin{figure*}
  \begin{center}
    \leavevmode
    \begin{center}
      \begin{tabular}{ll}
        \includegraphics[angle=-90,width=7.1cm]{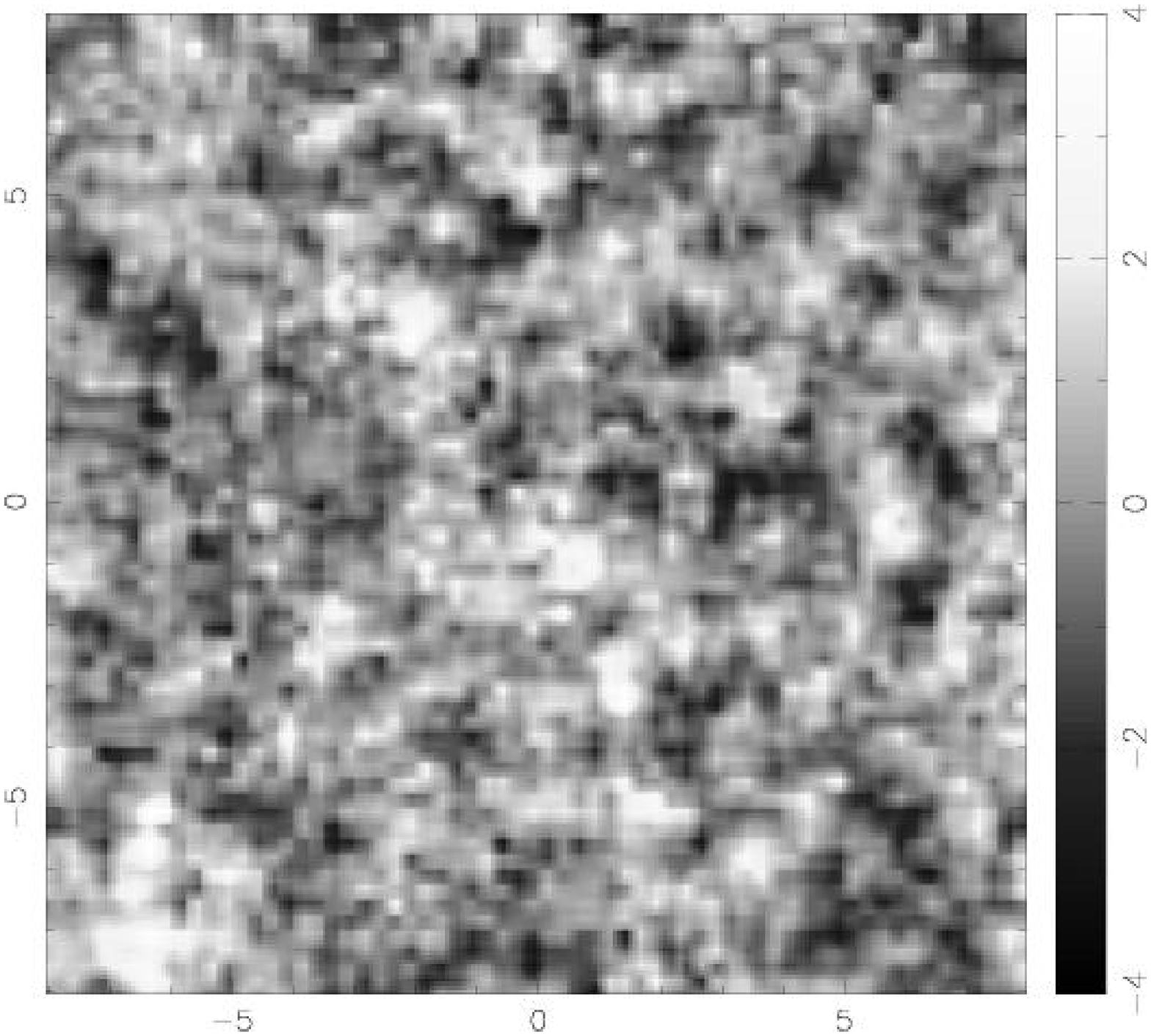}&
        \includegraphics[angle=-90,width=7.1cm]{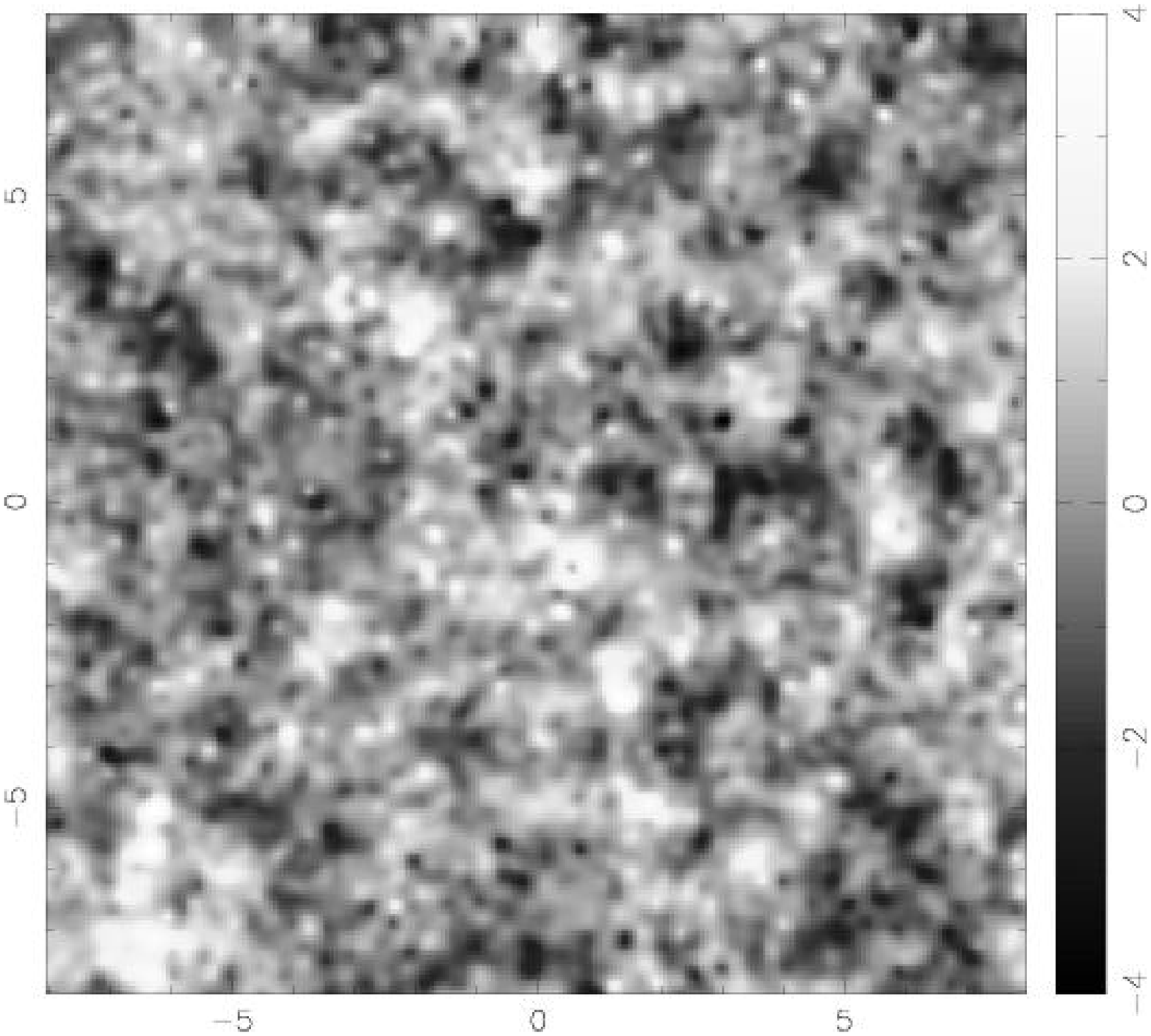}\\
        (a) & (b)\\
        \includegraphics[angle=-90,width=7.1cm]{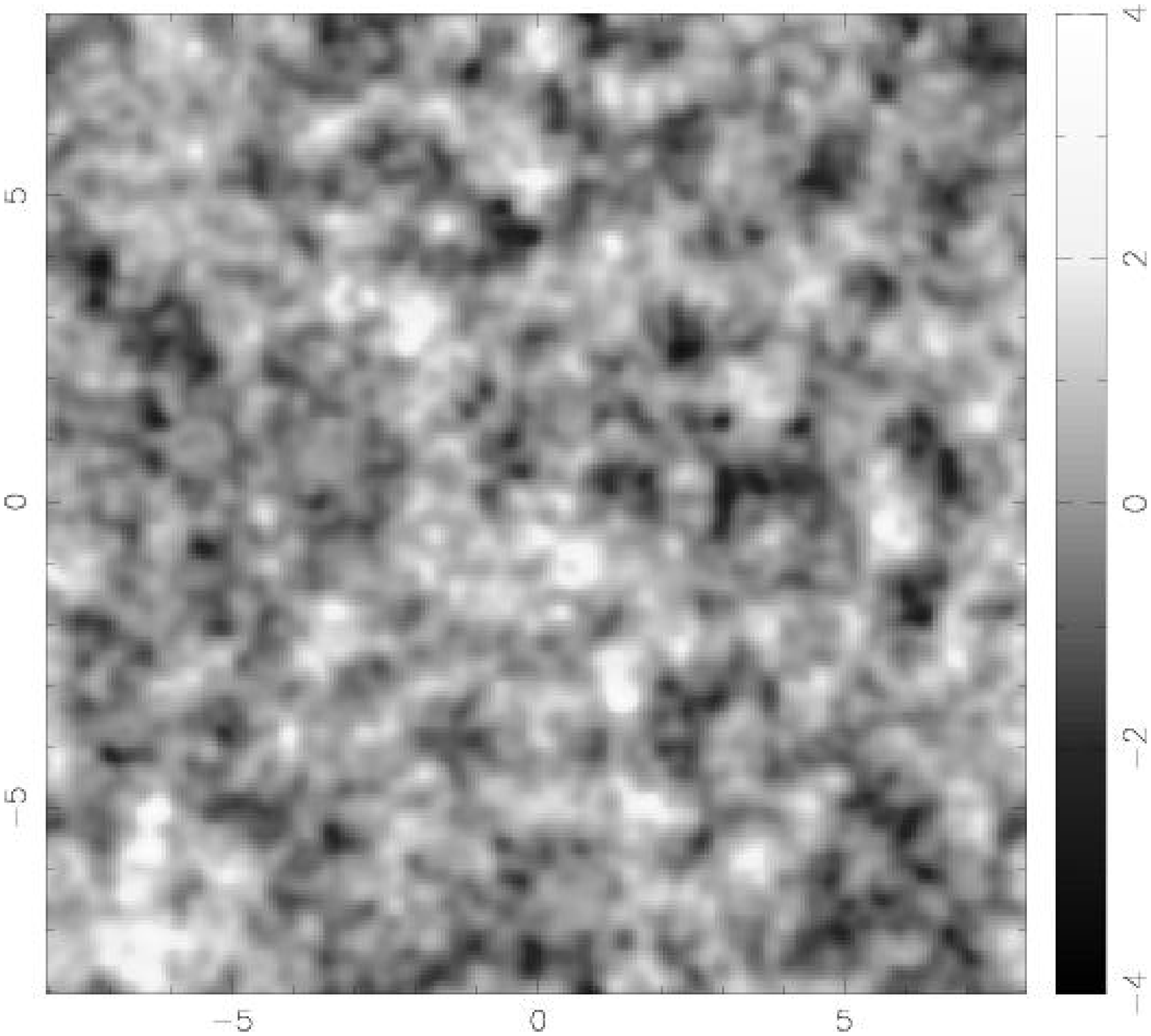}&
        \includegraphics[angle=-90,width=7.1cm]{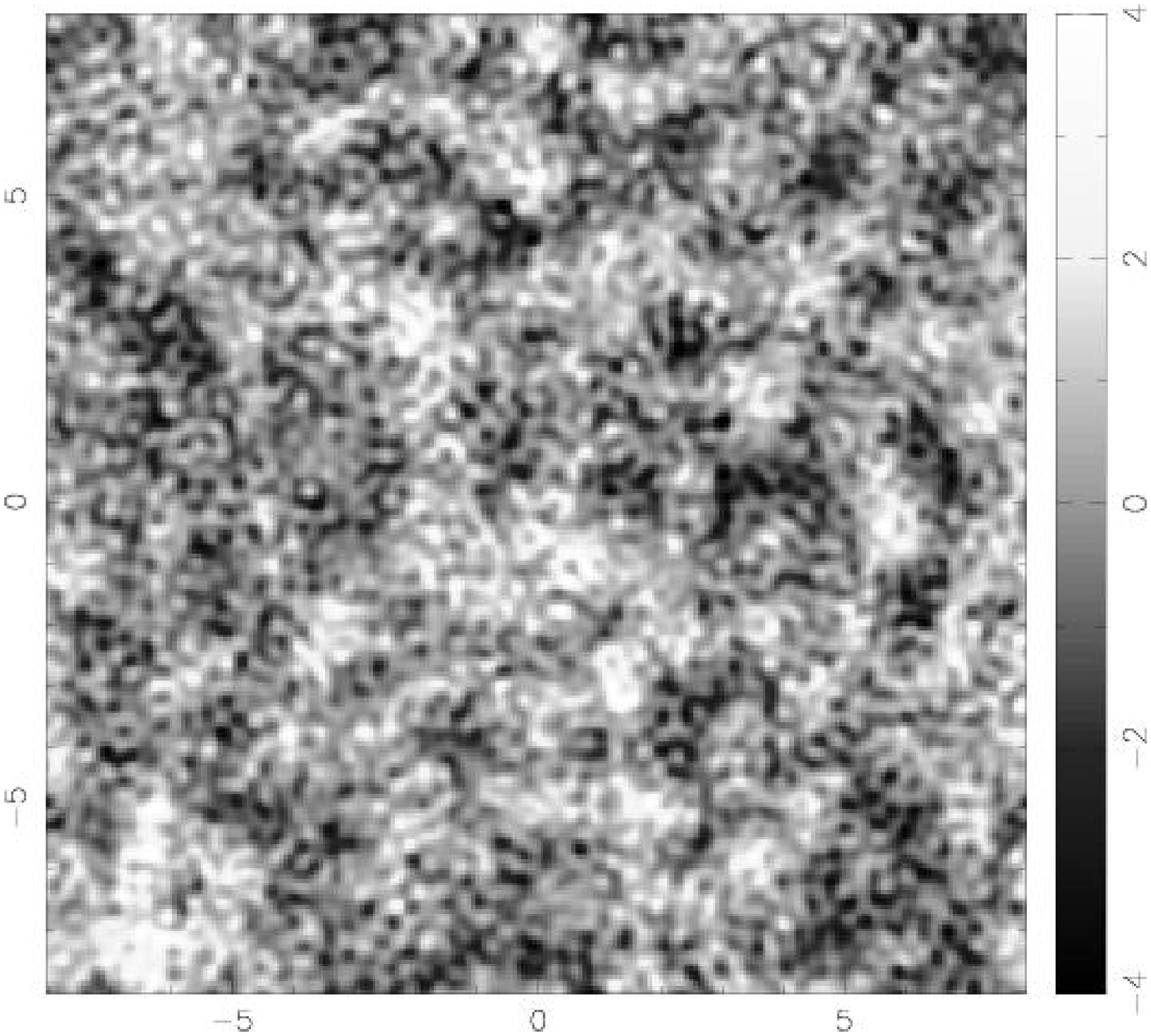}\\
        (c) & (d)
      \end{tabular}
    \end{center}
    \caption{Reconstructions of the data
      from~Fig.~\ref{fig:cmb:data}~(b) using the `wavelet MEM'
      algorithm: (a)~with Daubechies-4 tensor wavelets; (b) with
      4-level \`{a} trous wavelets.  A reconstruction with real-space
      historic MEM is shown in~(c), and a real-space reconstruction
      using classic MEM with a Bayesian choice of the regularisation
      constant in~(d).  Compare the original image in
      Fig.~\ref{fig:cmb:data}~(a).  }
    \label{fig:cmb:reg}
  \end{center}
\end{figure*}
For illustration, one of the realisations of the CMB maps used in the
simulations is shown in Fig.~\ref{fig:cmb:data}~(a). A simulated
observation of this map using a point spread function with a FWHM of
5~pixels and Gaussian random noise of 0.5 is shown in
Fig.~\ref{fig:cmb:data}~(b). Reconstructions using wavelet MEM are
shown in Figs.~\ref{fig:cmb:reg}~(a) for Daubechies-4 tensor wavelets
and~(b) for \`{a} trous wavelets. While the reconstruction errors are
virtually identical in both cases, the tensor reconstruction show
distinctly non-gaussian features introduced by the spiky wavelet
functions. Real-space MEM reconsructions of the same data are plotted
in Figs.~\ref{fig:cmb:reg}~(c) and~(d). The reconstruction~(c) was
obtained with the historic MEM criterion; reconstruction errors are
similar to those of the wavelet methods. The image~(d) was obtained
from a Bayesian choice of the regularisation constant with the classic
MEM. The reconstruction quality is visibly inferior, which is
confirmed by the reconstruction errors (0.59 compared to 0.40 for
historic MEM).

\begin{table*}
  \begin{center}
    \leavevmode
    \begin{tabular}{c||ccc|ccc|ccc}
      \hline \hline
      PSF {FWHM} & \multicolumn{3}{c|}{3} &\multicolumn{3}{c|}{5}
         &\multicolumn{3}{c}{10} \\
      noise \rms & 0.1 & 0.5 & 2.0 & 0.1 & 0.5 & 2.0 & 0.1 & 0.5 & 2.0 \\
         \hline 
         real-space MEM& 0 & 0 & 0 & 0 & 0 & 0 & 0 & 0 & 0\\[0.5ex] 
         tensor wavelet MEM&  8474 & 3437 & 930 & 2121 & 1287 & 431 & 155 & 268 & 90\\ 
         \`{a} trous wavelet MEM& 6197 & 2526 & 631 & 2305 & 1262 & 423 & 542 & 296 & 75\\
         \hline \hline
   \end{tabular}
    \caption{The logarithm~$\ln \Pr(\myvec{d})$ of the
      Bayesian evidence~$\Pr(\myvec{d})$ for differnt reconstructions
      of the CMB maps.  The values are averaged over a set of
      25~simulations with 5~different noise and image realisations and
      are normalised such that they equal zero for the real-space MEM
      for each dataset.}
    \label{tab:memwavelet:evidence:cmb}
  \end{center}
\end{table*}
The averaged logarithms~$\ln \Pr(\myvec{d})$ of the Bayesian
evidence~$\Pr(\myvec{d})$ obtained from reconstructions of the CMB
maps with the classic MEM criterion are shown in
Table~\ref{tab:memwavelet:evidence:cmb}.  They confirm the results
from Table~\ref{tab:memwavelet:evidence} and the visual impression
from Fig.~\ref{fig:cmb:reg}.  The evidences for the wavelet methods
are again significantly higher than for the real-space MEM, even
though the relative evidence ratios are not quite as pronounced as for
the reconstructions of the photographic image.


\section{Conclusions}
\label{sec:memwavelet:discussion}

Wavelets are functions that enable an efficient representation of
signals or images.  They help to identify and compress the information
content of the signal into a small number of parameters.  Because
wavelet functions span a whole range of spatial scales, they can be
used to describe signal correlations of different
characteristic lengths.  In this paper, we have investigated how
wavelets can be combined with the maximum entropy method to improve
the reconstruction of images from blurred and noisy data.

There are two principal ways to incorporate wavelets into the maximum
entropy method.  First, the wavelet transform can be treated as an
intrinsic correlation function that is used to decorrelate the data
(wavelet MEM).  Secondly, the wavelet transform can be combined with
the entropy functional into a new effective wavelet entropy (wavelet
regularised entropy).  We have implemented both approaches for
orthogonal wavelet transforms.  Another type of wavelet transform is
the \`{a} trous transform.  It is non-orthogonal and has the benefit
of being invariant under translations and rotations of the image.  We
show that the \`{a} trous transform can be considered as a special
case of a multi-channel ICF, in which the image is produced from a
linear combination of images convolved with point spread functions of
different widths.  The quality of the reconstruction depends on the
relative weights assigned to each channel or scale.

We have applied MEM implementations using both orthogonal and \`{a}
trous wavelets to simulated observations of CMB temperature
anisotropies.  We find that while the relative weighting of scales or
channels is important, there is a range of different weightings that
can yield roughly similar results as long as they suppress small-scale
structure in the image.  It does not matter much whether the weighting
is introduced by setting different channel weights or by rescaling the
entropy expressions or default models.  Weightings that suppress
small-scale structure more efficiently can perform better for low
signal-to-noise, while they are worse for high signal-to-noise.
Furthermore, we also find that for images containing structure on 
different scales, like CMB maps, methods that try to improve
the reconstruction by ad hoc assignments of pixel- and data-dependent
weights usually do not enhance the reconstruction quality.  The more
complicated reconstruction prescriptions given by
\citet{pantinstarck96} have no benefits for CMB map-making.

As far as reconstruction errors are concerned, wavelet-based maximum
entropy algorithms seem to match the standard MEM in pixel space
(real-space MEM) for large point spread
functions or low noise levels.  There exist sufficient well-determined
degrees of freedom in the data to make the image basis irrelevant.  On
the other hand, for poor signal-to-noise and narrow convolution masks,
wavelet methods outperform real-space MEM.  Thus the use of wavelet
techniques can improve the reconstruction of images in many cases,
while there is no disadvantage of using these methods in other
situations.  The improvement seems to be genuinely related to the
basis set or ICF and not just an artefact of an improper choice of the
regularisation.  A Bayesian treatment of the regularisation constant
and a comparison of the different reconstruction methods shows a much
higher evidence for ICF methods than for the simple real-space MEM.
In a Bayesian context, the wavelet basis can thus be interpreted as a
better `model' for the image.  In this regard, it fulfills the promise
of an improvement over the real-space method that the ICF was designed
to address \citep[see e.g.][]{mackay92a}.

The isotropic \`{a} trous transform or the multi-channel ICF can in
some cases improve on orthogonal wavelets.  Furthermore, they can be
implemented within the \textsc{memsys5} maximum entropy kernel and
thus be applied in a proper Bayesian maximisation scheme.  To
summarise, we conclude that the use of wavelets in MEM image
reconstructions is a successful technique that can improve the quality
of image reconstructions.

\section*{Acknowledgments}

We thank Belen Barreiro, Steve Gull, Phil Marshall and Charles McLachlan for
helpful discussions.  
Klaus Maisinger acknowledges support from an EU Marie Curie
Fellowship.



\appendix

\section{Multiresolution Analysis}
\label{sec:wavelet:mra}

Multiresolution analysis provides a simple framework for describing the
properties of wavelet and scaling functions discussed in 
Section~\ref{sec:memwavelet:wavelets}. 
Let~$\mathcal{L}^2(\mathset{R})$ be the set of square integrable
functions.  The multiresolution analysis is a sequence of closed
subspaces $\{ \mathcal{V}_j\}_{j \in \mathset{Z}} \subset
\mathcal{L}^2(\mathset{R})$ which
approximate~$\mathcal{L}^2(\mathset{R})$.  The subspaces are nested
\[
 \ldots \subset \mathcal{V}_{-1} \subset \mathcal{V}_0 \subset \mathcal{V}_1 \subset \ldots
\]
such that their union $\bigcup_{j \in \mathset{Z}} \mathcal{V}_j$ is
dense in~$\mathcal{L}^2(\mathset{R})$, and for any function~$f(x)$
\[ 
f(x) \in \mathcal{V}_j \Leftrightarrow f(2x) \in \mathcal{V}_{j+1}.
\]
In this picture, the scaling functions~$\{\phi(x-l), l \in \mathset{Z}
\}$ form an orthonormal basis in the reference space~$\mathcal{V}_0$.
Because $\mathcal{V}_0 \subset \mathcal{V}_1$, elements of
$\mathcal{V}_0$ can be written as linear combinations of those of
$\mathcal{V}_1$:
\begin{equation}
\phi(x) = \sum_k \sqrt{2} H_k \phi(2x-k).\label{eqn:refinement}
\end{equation}
This expression is called a refinement relation for the scaling
function~$\phi(x)$.

Given a subspace~$\mathcal{V}_j$ and its basis of scaling
functions~$\phi(x)$, one can ask how this basis in~$\mathcal{V}_j$ can
be extended to a basis in~$\mathcal{V}_{j+1}$.  The wavelet
functions~$\psi(x)$ can be introduced as a set of functions needed to
complete the basis in~$\mathcal{V}_{j+1}$.  In other words, they form
a basis of the orthogonal complement $\mathcal{W}_j$ of
$\mathcal{V}_j$, i.e. $\mathcal{V}_{j+1} = \mathcal{V}_j \oplus
\mathcal{W}_j$.  Consequently, the wavelet functions can be written in
the form of a refinement relation
\begin{equation}
\psi(x) = \sum_k \sqrt{2} G_k \phi(2x-k).
\end{equation}
The coefficients $G_k$ are related to the $H_k$ via $ G_k = (-1)^k
H_{1-k}$.  In a signal processing context, the sequences $H_k$ and
$G_k$ are called {\em quadrature mirror filters}. The sequence $H_k$
is a low pass filter, while $G_k$ acts as a high pass filter.

The wavelet transform represents a function in terms of its smooth
basis functions $\phi_{0,l} \in \mathcal{V}_0$ and the detail
functions $\psi_{j,l} \in \mathcal{W}_j$.
 
\subsection{The discrete wavelet transform} 
\label{sec:appendix:dwt}

For the discrete wavelet transform introduced in
Section~\ref{sec:wavelet:dwt}, wavelet coefficients can be constructed
directly from the data vector $c_{J,l} = f(x_l)$ ($l = 1,\ldots,N$)
via
\begin{eqnarray}
w_{j-1, k} &=& \sum_l H_{l-2k} c_{j,l},\nonumber \\
c_{j-1, k} &=& \sum_l G_{l-2k} c_{j,l}.\label{eqn:dwtfiltercoeff}
\end{eqnarray}
This prescription leads to a pyramidal algorithm for the
implementation of the discrete wavelet transform. At each iteration,
the data vector of length $2^j$ is split into $2^{j-1}$ detail
components and $2^{j-1}$ smoothed components. The smoothed components
are then used as input for the next iteration to reconstruct the
detail coefficients at the next larger scale.

\subsection{The \`{a} trous transform} 
\label{sec:appendix:atrous}

For the \`{a} trous transform from Section~\ref{sec:wavelet:atrous}, a
data vector $c_{J,l}$ ($l=1,\ldots,N$, $2^J \le N$) is iteratively
smoothed with a filter mask~$H_l$ (\ref{eqn:atrous}):
\[
c_{j-1,k} = \sum_l H_l c_{j,k+2^{(J-j)} l}.
\]
The coefficients~$H_l$ can be derived from a refinement relation of
the form
\[
\phi(x) = \sum_l 2 H_l \phi (2x-l),
\]
where~$\phi(x)$ is a scaling function.  Note the difference of a
factor~$\sqrt{2}$ compared to~(\ref{eqn:refinement}).  
The  \`{a} trous wavelet vector is given by
\[
w_{j-1,k} = \sum_l G_l c_{j,k+2^{(J-j)} l} = c_{j,k} - c_{j-1,k},
\]
from which $G_l = \delta_{l,0} - H_l$.  For the wavelet~$\psi (x)$
one finds the scaling relation
\[
\psi (x) = \sum_l 2 G_l \phi (2x-l) = 2 \phi(2x) - \phi (x).
\]

\section{Two-dimensional wavelet transforms}
\label{sec:appendix:transform2d}

\subsection{Tensor products}

\begin{figure}
  \begin{center}
    \leavevmode
    \includegraphics[width=5cm]{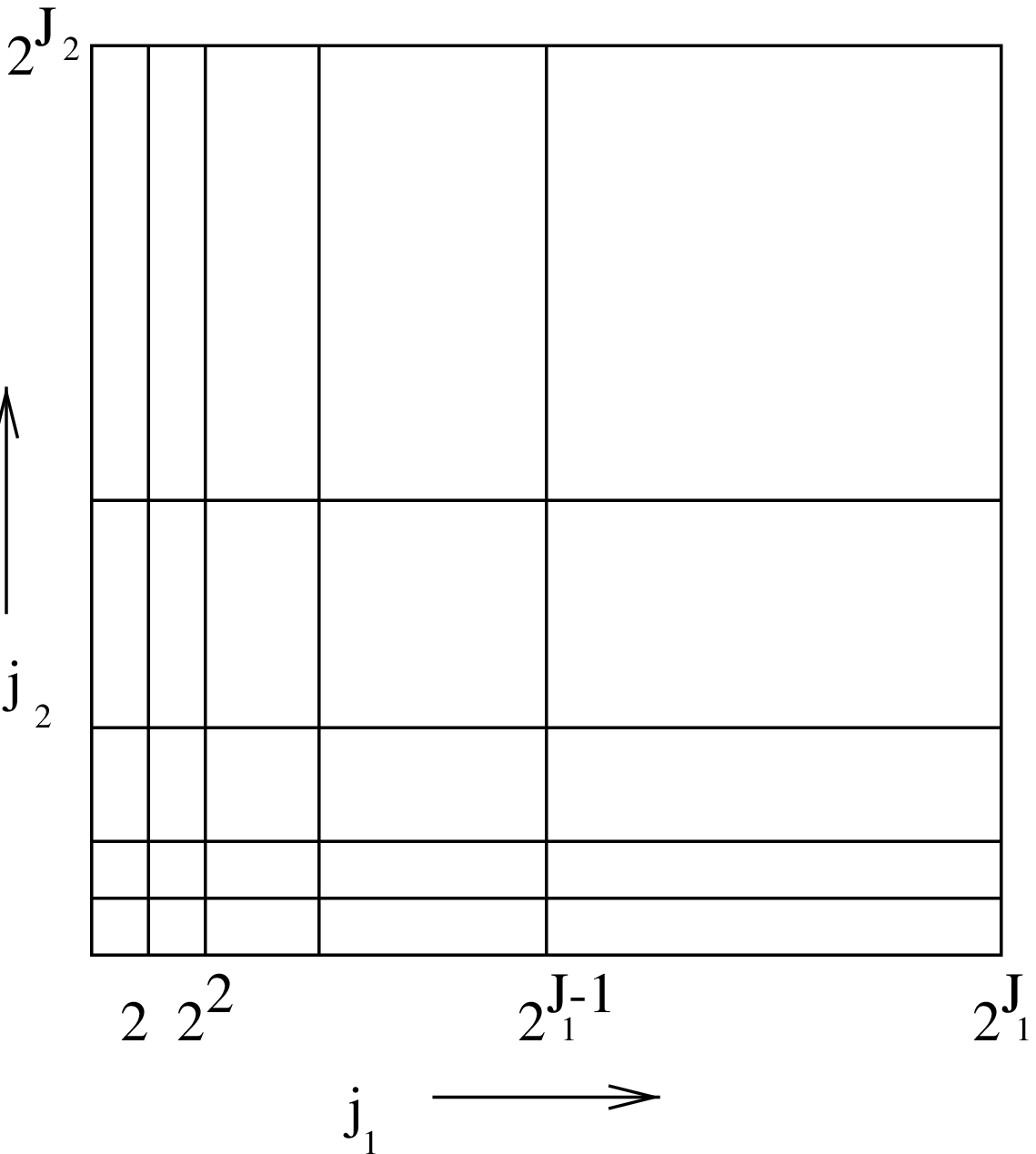}
    \caption{The partitioning of the wavelet image into different
      domains  for the tensor transform. There are $J \times J$
      domains mixing different scales $j_1$ and $j_2$. Increasing $j$
      corresponds to finer scales.} 
    \label{fig:tensordomains}
  \end{center}
\end{figure}
The simplest approach to extending the wavelet transform to two
dimensions uses tensor products of the one-dimensional wavelets as a
two-dimensional basis \citep[e.g.][]{press92}. The resulting algorithm
is straightforward. First one performs a wavelet transform on the
first index of the image matrix for all possible values of the second,
and then on the second.  The tensor basis functions mix
one-dimensional wavelets from different scales; they are given by
\begin{eqnarray*}
\phi_{0,0;l_1,l_2}(x,y) & = & \phi_{0,l_1}(x)\phi_{0,l_2}(y), \\
\zeta_{j_1,0;l_1,l_2}(x,y) & = & \psi_{j_1,l_1}(x)\phi_{0,l_2}(y), \\
\xi_{0,j_2;l_1,l_2}(x,y) & = & \phi_{0,l_1}(x)\psi_{j_2,l_2}(y), \\
\psi_{j_1,j_2;l_1,l_2}(x,y) & = & \psi_{j_1,l_1}(x)\psi_{j_2,l_2}(y).
\end{eqnarray*}
The two-dimensional pixelised image has dimensions $N \times N =
2^{J_1} \times 2^{J_2}$ and by analogy with~(\ref{eqn:dwtexpansion})
one has $0 \leq j_1 \leq J_1-1$ and $0 \leq l_1 \leq 2^{j_1}-1$, and
similarly for $j_2$ and $l_2$.  If we denote such an image by the ($N
\times N$-) matrix~$\mymatrix{T}$, then the matrix of wavelet
coefficients is given by
\begin{equation}
\widetilde{\mymatrix{T}} = \mymatrix{W}^{(2d)} \mymatrix{T} = \mymatrix{W}^{(1d)}\  \mymatrix{T}\
  \mymatrix{W}^{(1d) \transp},
\label{eqn:dwt2d}
\end{equation}
where $\mymatrix{W}^{(1d)}$ is the $N \times N$-matrix describing the
one-dimensional transform that was introduced in~(\ref{eqn:dwt1d}),
and $\mymatrix{W}^{(1d) \transp}$ is its transpose.  The
matrix~$\widetilde{\mymatrix{T}}$ is partitioned into $J_1 \times J_2$
separate domains of $2^{j_1} \times 2^{j_2}$ wavelet coefficients
(see Fig.~\ref{fig:tensordomains}), according to the scale indices
$j_1$ and $j_2$ in the horizontal and vertical directions
respectively.  By analogy with the one-dimensional case, as $j_1$
increases the wavelets represent the horizontal structure in the image
on increasingly smaller scales. Similarly, as $j_2$ increases the
wavelets represent the increasingly fine scale vertical structure in
the image.  Thus domains that lie in the leading diagonal (i.e. with
$j_1=j_2$) contain coefficients of two-dimensional wavelets that
represent the image at the same scale in the horizontal and vertical
directions, whereas domains with $j_1 \neq j_2$ contain coefficients
of two-dimensional wavelets describing the image on different scales
in the two directions.

\subsection{MRA transforms}

\begin{figure}
  \begin{center}
    \leavevmode
    \includegraphics[width=5cm]{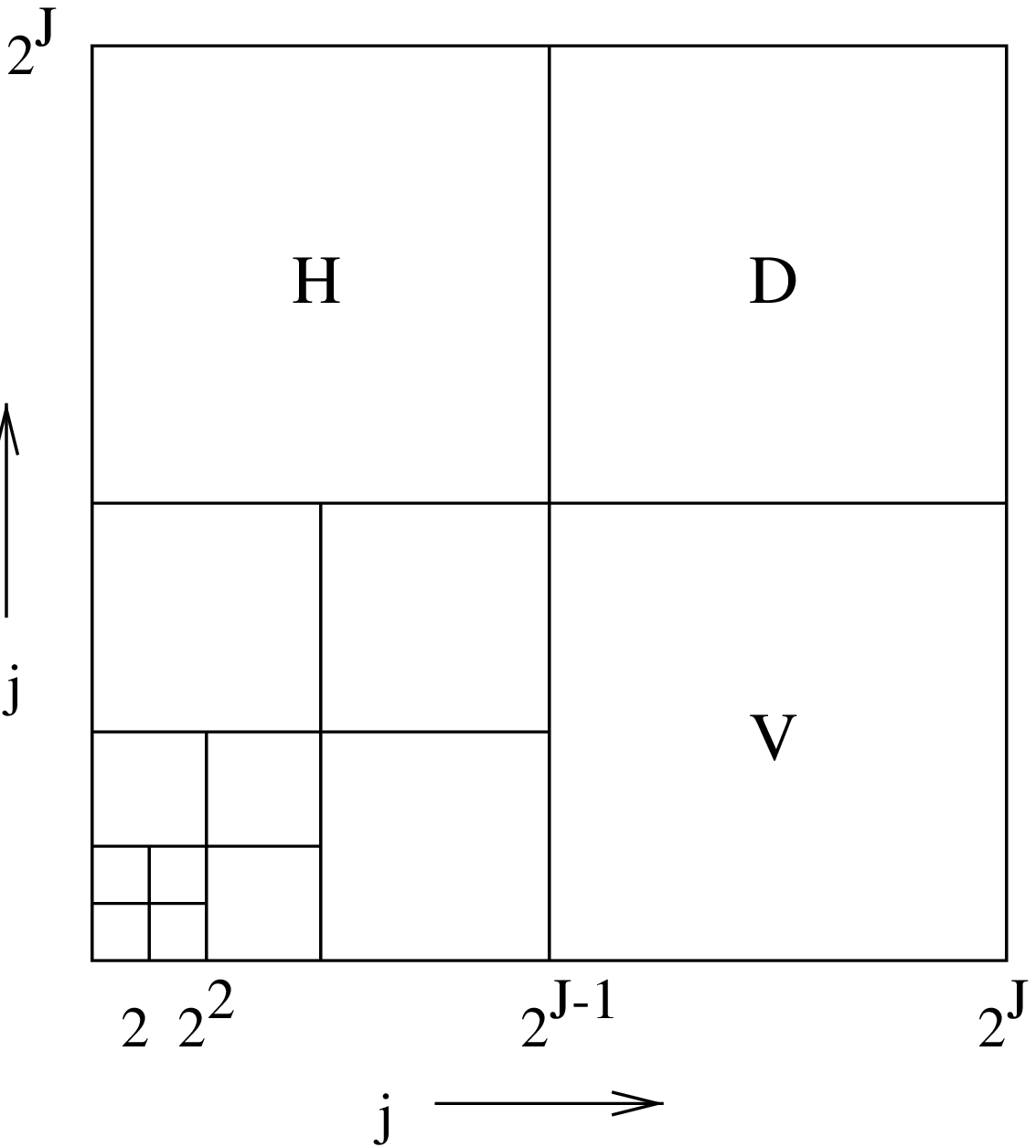}
    \caption{The partitioning of the wavelet image into different
      domains for the MRA transform. At each scale $j$, there are
      3~different domains: horizontal, vertical and diagonal. Again,
      increasing $j$ corresponds to finer scales.}
    \label{fig:mradomains}
  \end{center}
\end{figure}
The MRA transform uses three different tensor
products of one--dimensional wavelets.  Dilated versions of these
tensor products form the detail wavelets.  They do not mix scales and
can be described in terms of a single scale index $j$.  At each scale
level $j$, these bases are given by
\begin{eqnarray*}
\phi_{0;l_1,l_2}(x,y) & = & \phi_{0,l_1}(x)\phi_{0,l_2}(y), \\
\psi^\mathrm{H}_{j;l_1,l_2}(x,y) & = & \psi_{j,l_1}(x)\phi_{j,l_2}(y), \\
\psi^\mathrm{V}_{j;l_1,l_2}(x,y) & = & \phi_{j,l_1}(x)\psi_{j,l_2}(y), \\
\psi^\mathrm{D}_{j;l_1,l_2}(x,y) & = & \psi_{j,l_1}(x)\psi_{j,l_2}(y).
\end{eqnarray*}
The $\phi_{j;l_1,l_2}$ wavelet is simply an averaging function at the
$j$th level, while the other three wavelets correspond to structure at
the $j$th scale level in the horizontal, vertical and diagonal
directions in the image. The structure of the matrix
$\widetilde{\mymatrix{T}}$ from~(\ref{eqn:dwt2d}) is different for the
MRA transform.  Obviously, the matrix is square and its domains can be
described by a single value of $J$. Fig.~\ref{fig:mradomains} shows
how it is partitioned into $3(J-1)$~detail domains and one domain
with the smoothed components.

\section{Calculation of derivatives}
\label{sec:appendix:memwavelet}

In an implementation of the maximum entropy algorithm, the posterior
functional $F (\myvec{h})= \frac{1}{2} \chi^2 (\myvec{h}) - \alpha S
(\myvec{h})$ has to be minimised numerically.  Some minimisation
routines, like the \textsc{memsys5} package \citep{gullskilling99},
require first derivatives of $F$ with respect to the (visible or
hidden) image pixels, while others, like the simple Newton-Raphson
method, additionally require second derivatives.  The derivatives of the
$\chi^2$- and entropy terms can be calculated separately.
The $\chi^2$-functional
for a linear instrumental response matrix~$\mymatrix{R}$ is given by
\[
\chi^2(\myvec{h}) = (\mymatrix{R} \mymatrix{K} \myvec{h} - \myvec{d})^\transp
\mymatrix{N}^{-1} (\mymatrix{R} \mymatrix{K} \myvec{h} - \myvec{d}),
\]
where $\myvec{d}$ is the data vector and $\mymatrix{K}$ the ICF.  The
gradient and curvature can be derived straightforwardly:
\begin{eqnarray}
\nabla_{\myvec{h}} \chi^2& = &(\mymatrix{R} \mymatrix{K})^\transp
\mymatrix{N}^{-1} (\mymatrix{R} \mymatrix{K} \myvec{h} - \myvec{d}),
\label{eqn:chi2gengrad} \\ 
\nabla_{\myvec{h}}\nabla_{\myvec{h}} \chi^2& = &\mymatrix{K}^\transp
\mymatrix{R}^\transp \mymatrix{N}^{-1} \mymatrix{R} \mymatrix{K}.
\label{eqn:chi2gencurv}
\end{eqnarray}
For an implementation, one thus needs to provide the transformation
$\mymatrix{R}$, the wavelet transform
$\mymatrix{W}=\mymatrix{K}^\transp$ and their
transposes~$\mymatrix{R}^\transp$ and~$\mymatrix{W}^\transp$.

\subsection{Real-space MEM}
\label{sec:appendix:memwavelet:realmem}

For a maximum entropy algorithm operating on data and reconstructed
images given in real space, i.e. in the same image plane, the data can
be predicted by a convolution of the underlying image
distribution~$h(\myvec{x})$ with a point spread
function~$P(\myvec{x})$.  The discrete predicted data
samples~$d_i^{\rm P}$ are given by
\begin{equation}
d_i^{\rm P} = \sum_j R_{ij} h(\myvec{x}_j) = \sum_j
P(\myvec{x}_i-\myvec{x}_j) h(\myvec{x}_j).
\label{eqn:traforeal}
\end{equation}
The transpose~$\mymatrix{R}^\transp$ of the response matrix is easily
obtained. Its operation on an image vector $\myvec{h}$, where $h_i =
h(\myvec{x}_i)$ is given by
\begin{equation}
\sum_j P (-\myvec{x}_i+\myvec{x}_j) h(\myvec{x}_j).
\label{eqn:traforealtranspose}
\end{equation}
This is a correlation of the functions~$P(\myvec{x})$
and~$h(\myvec{x})$, and not a convolution.  For a symmetric beam,
$\mymatrix{R} = \mymatrix{R}^\transp$.  The terms $R_{ij}^2$, which
are useful to derive the curvature~(\ref{eqn:chi2gencurv}), are also
straightforward to calculate.  The convolution can be most easily
implemented with FFTs.  By substituting~(\ref{eqn:traforeal})
and~(\ref{eqn:traforealtranspose}) into the
expressions~(\ref{eqn:chi2gengrad}) and~(\ref{eqn:chi2gencurv}), and
setting the ICF to the identity~$\mymatrix{K}=\mymatrix{1}$, one can
obtain the derivatives of the $\chi^2$-functional.

\subsection {Wavelet MEM}
\label{sec:appendix:memwavelet:waveletmem}

Wavelet MEM uses the real-space response matrix derived in
Section~\ref{sec:appendix:memwavelet:realmem}, and an
ICF~$\mymatrix{K}$ that is given by the
transpose~$\mymatrix{W}^\transp$ of the wavelet transform.
Derivatives can by obtained by substitution into the
expressions~(\ref{eqn:chi2gengrad}) and~(\ref{eqn:chi2gencurv}).  For
the orthogonal wavelet transforms, the transpose of the wavelet
transform is simply the inverse transform.  The transpose of the \`{a}
trous transform is given by~(\ref{eqn:atroustransp}).

We note that the $\chi^2$-curvature
\[ 
\nabla_{\myvec{h}} \nabla_{\myvec{h}} \chi^2 = \mymatrix{W}
\mymatrix{R}^\transp \mymatrix{N}^{-1} \mymatrix{R}
\mymatrix{W}^\transp
\]
is numerically difficult to evaluate efficiently for the orthogonal
transforms.  However, for a diagonal noise covariance~$\mymatrix{N}$,
$N_{ii} = \sigma_i^2$, the diagonal elements of the curvature are
\[
\sspd{\chi^2}{h_n} = \sum_{i,k,o,p} W_{nk} R_{ki}^\transp
\frac{2}{\sigma_i^2} R_{io} W_{op}^\transp \delta_{pn}.
\]
If the noise covariance is constant across pixels with
$\sigma_i=\sigma$, this reduces to 
\[
\sspd{\chi^2}{h_n} = \sum_{ikop} \frac{2}{\sigma^2} W_{nk}
R_{ki}^\transp R_{io} W_{op}^\transp \delta_{pn},
\]
which only needs to be evaluated once for each wavelet domain.  This
makes a calculation of the curvature feasible in a reasonable amount
of time.

\subsection{Wavelet-regularised MEM}
\label{sec:appendix:memwavelet:waveregmem}

In wavelet-regularised MEM, the data~$\myvec{d}^{\rm P}$ are predicted
from an image~$\myvec{h}$ in the same way as in the real-space
algorithm from Section~\ref{sec:appendix:memwavelet:realmem}.  The
entropy functional, however, is calculated on the wavelet coefficients
rather than the image pixels.  This can be viewed as a new entropy
functional created from a composition of the standard entropy and the
wavelet transform.  Numerical derivatives are given by  
\begin{eqnarray*}
\nabla_{\myvec{v}} [S(\mymatrix{W} \myvec{v})]  &=& 
\mymatrix{W}^\transp [\nabla_{\myvec{h}} S] (\mymatrix{W} \myvec{v}),
\\
\nabla_{\myvec{v}}\nabla_{\myvec{v}} [S(\mymatrix{W} \myvec{v})] &=& 
\mymatrix{W}^\transp [\nabla_{\myvec{h}} \nabla_{\myvec{h}} S] (\mymatrix{W} \myvec{v}) \mymatrix{W}.
\end{eqnarray*}
Again, second derivatives are numerically difficult to evaluate for
orthogonal wavelet transforms.

\section{Equivalence of methods}
\label{sec:wavemem:equivalence}

In this section, we show that the `wavelet MEM' and `wavelet regularised
MEM' introduced in Section~\ref{sec:memwavelet:transforms} become
equivalent if the regularisation function is quadratic and the wavelet
transform orthogonal.

For wavelet MEM, the $\chi^2$-statistic is given by
\begin{equation}
\chi^2(\myvec{h}) = (\mymatrix{R}\mymatrix{K}\myvec{h}
-\myvec{d})^\transp \mymatrix{N}^{-1} (\mymatrix{R}\mymatrix{K}\myvec{h}
-\myvec{d}).
\end{equation}
The most general form of quadratic regularisation is given by
\[
S(\myvec{h})  = - \myvec{h}^\transp
\mymatrix{M}^{-1} \myvec{h}.
\]
In the special case of the quadratic
approximation~(\ref{eqn:quadentropy}) to the positive/negative
entropy with a model~$\myvec{m}$, the matrix~$\mymatrix{M}$ is defined
as $M_{ij} = 4 m_i \delta_{ij}$.  The maximum entropy
solution~$\hat{\myvec{h}}$ can be found by minimising the
function~(\ref{eqn:imagefunc}),
\begin{eqnarray*}
F(\myvec{h}) & = & \tfrac{1}{2} \chi^2(\myvec{h}) - \alpha S(\myvec{h})\\
&=& \tfrac{1}{2} (\mymatrix{R}\mymatrix{K}\myvec{h}
-\myvec{d})^\transp \mymatrix{N}^{-1} (\mymatrix{R}\mymatrix{K}\myvec{h}
-\myvec{d}) + \alpha \myvec{h}^\transp
\mymatrix{M}^{-1} \myvec{h}\\
&=& \tfrac{1}{2} \myvec{d}^\transp \mymatrix{N}^{-1} \myvec{d} -
\myvec{h}^\transp \mymatrix{K}^\transp \mymatrix{R}^\transp
\mymatrix{N}^{-1} \myvec{d}\\
&& + \myvec{h}^\transp (\tfrac{1}{2} \mymatrix{K}^\transp \mymatrix{R}^\transp
\mymatrix{N}^{-1} \mymatrix{R}\mymatrix{K} + \alpha \mymatrix{M}^{-1})
\myvec{h}. 
\end{eqnarray*}
Demanding that the gradient of~$F$ with respect to~$\myvec{h}$ vanishes
at the maximum~$\hat{\myvec{h}}$, we obtain 
\begin{equation}
(\tfrac{1}{2} \mymatrix{K}^\transp \mymatrix{R}^\transp
\mymatrix{N}^{-1} \mymatrix{R}\mymatrix{K} + \alpha \mymatrix{M}^{-1})
\hat{\myvec{h}} = \mymatrix{K}^\transp \mymatrix{R}^\transp
\mymatrix{N}^{-1} \myvec{d}.
\label{eqn:solwavmem}
\end{equation}

For wavelet regularised MEM, we have
\begin{eqnarray*}
\chi^2(\myvec{v}) &= &(\mymatrix{R}\myvec{v}
-\myvec{d})^\transp \mymatrix{N}^{-1} (\mymatrix{R}\myvec{v}
-\myvec{d}) \quad \mathrm{and}\\
S(\mymatrix{K}^\transp \myvec{v})& =& - \myvec{v}\mymatrix{K}
\mymatrix{M}^{-1}\mymatrix{K}^\transp \myvec{v}. 
\end{eqnarray*}
The maximum entropy solution $\hat{\myvec{v}}$ is given by
\begin{equation}
(\tfrac{1}{2} \mymatrix{R}^\transp
\mymatrix{N}^{-1} \mymatrix{R} + \alpha \mymatrix{K}\mymatrix{M}^{-1} \mymatrix{K}^\transp )
\hat{\myvec{v}} = \mymatrix{R}^\transp
\mymatrix{N}^{-1} \myvec{d}.
\label{eqn:solwavregmem}
\end{equation}
For orthogonal wavelet transforms $\mymatrix{W} =
\mymatrix{K}^\transp$, 
$
\mymatrix{K} \mymatrix{K}^\transp = \mymatrix{1} =\mymatrix{K}^\transp\mymatrix{K}.
$
Multiplying~(\ref{eqn:solwavregmem}) by~$\mymatrix{K}^\transp$, 
we obtain
\[
(\tfrac{1}{2} \mymatrix{K}^\transp\mymatrix{R}^\transp
\mymatrix{N}^{-1} \mymatrix{R}\mymatrix{K} + \alpha \mymatrix{M}^{-1} )
\mymatrix{K}^\transp \hat{\myvec{v}} = \mymatrix{K}^\transp\mymatrix{R}^\transp
\mymatrix{N}^{-1} \myvec{d}.
\]
By comparison with~(\ref{eqn:solwavmem}), we see that
$
\hat{\myvec{h}} =\mymatrix{K}^\transp \hat{\myvec{v}}. 
$
The solutions for wavelet MEM and wavelet regularised MEM are identical.


\label{lastpage}

\end{document}